\documentclass[twocol]{ametsocV6.1}

\usepackage{placeins}  
\usepackage{siunitx}
\usepackage{csquotes}  

\title{Multi-model ocean oxygen fields predicted by conditional diffusion models}

\authors{Linus Vogt\aff{a}\correspondingauthor{Linus Vogt, linus.vogt@nyu.edu} 
and Laure Zanna\aff{a}
}

\affiliation{\aff{a}{Courant Institute School of Mathematics, Computing, and Data Science, New York University,
New York, NY, USA}}

\abstract{
Dissolved oxygen is important for the ocean's ecosystems and biogeochemical cycles.
Yet, Earth System Models (ESMs) vary in their simulations of the present-day and future ocean oxygen inventory.
To narrow down the uncertainty in estimates of ocean oxygen content, we train a conditional generative diffusion model on outputs of a multi-model ESM ensemble to learn the conditional distribution of upper-ocean oxygen given physical input variables such as temperature and salinity.
This generative model has considerable skill in the Atlantic and Southern Oceans, and can generate realistic oxygen samples under conditions not seen in the training data.
We validate this model using observational datasets, and use it to generate oxygen fields for models without oxygen data using only temperature and salinity as conditioning inputs.
This physics-conditioned extrapolation suggests that model biases in the tropical Pacific Oxygen Minimum Zone may be smaller than currently assumed when considering a larger set of physical ocean states.
Our approach provides a complementary way to represent probabilistic multi-model climate distributions.
}

\begin{document}

\maketitle

\FloatBarrier
\section{Introduction}

Dissolved oxygen (O$_2$) is a crucial component of marine ecosystem functioning and the regulation of marine biogeochemical cycles \citep{deutsch_climate_2024}.
As a result of anthropogenic climate change, the global ocean has experienced a decrease in its dissolved oxygen inventory of over 2\% since 1960 \citep{schmidtko_decline_2017,breitburg_declining_2018,ito_are_2026}.
This ocean deoxygenation occurs due to a combination of (i) reduced oxygen solubility in warming water masses and (ii) changes in ocean circulation and stratification as well as biological processes such as microbial respiration \citep{shepherd_ocean_2017,oschlies_drivers_2018,sallee_summertime_2021}.
As a result, the metabolism of marine animals is negatively affected \citep{deutsch_climate_2015}, and ecosystems are altered through habitat compression and range shifts of species \citep{limburg_ocean_2020}, often in combination with other stressors \citep{bopp_multiple_2013,boyd_biological_2015}.
In addition to this long-term trend, the increasing occurrence of low-oxygen extreme events places additional stress on ecosystems \citep{gruber_biogeochemical_2021,sampaio_impacts_2021,legrix_surface_2025}.

Simulations using coupled Earth System Models (ESMs) participating in the Coupled Model Intercomparison Project (CMIP) consistently predict a continuing ocean deoxygenation under future emissions scenarios \citep{keeling_ocean_2010,bopp_multiple_2013,long_oxygen_2019,kwiatkowski_twentyfirst_2020,oschlies_committed_2021}.
Although models agree on solubility-driven deoxygenation, simulations by different ESMs show considerable spread in the magnitude and pattern of the oxygen response to the same forcing, contributing to uncertainty in these projections \citep{frolicher_sources_2016,li_increasing_2020,takano_simulations_2023}.
In addition, ESMs tend to underestimate both the long-term trend \citep{ito_are_2026} as well as the temporal variability \citep{frolicher_natural_2009} in regional oxygen over the historical period.
These model biases and uncertainties arise from the insufficient representation of relevant physical and biogeochemical processes affecting the ocean's oxygen budget \citep{stramma_mismatch_2012,andrews_biogeochemical_2017,oschlies_patterns_2017,palter_sensitivity_2018,buchanan_regional_2021}, especially in tropical Oxygen Minimum Zones \citep[OMZs;][]{busecke_diverging_2022,ditkovsky_unifying_2025,shigemitsu_one_2017,delteil_emerging_2026}.

Recent advances in data-driven and machine learning (ML) techniques offer novel ways of constraining ocean deoxygenation.
Although direct oxygen measurements are relatively sparse in time and space, supervised ML approaches have been successfully used to produce gridded datasets of oxygen concentration and other tracers over recent decades \citep[e.g.,][]{keppler_recent_2023,sharp_gobaio2_2023,ito_mapping_2024,lu_oxygenerator_2024,piedagnel_mapping_2025,zhang_machine_2026}.
These approaches are generally based on learning the relationship between oxygen and a number of physical predictor variables, such as temperature and salinity, which are more readily available from shipboard and float observations.

A particular class of ML models that has shown promising performance on ocean and climate datasets is the Denoising Diffusion Probabilistic Model \citep[DDPM;][]{sohl-dickstein_deep_2015,ho_denoising_2020,nichol_improved_2021}.
In broad terms, the generic task of these models is to generate samples from the distribution $p_\mathrm{data}(\mathbf{x})$ of a high-dimensional dataset, such as a set of images or spatial climate fields $\mathbf{x}$.
In practice, this is commonly achieved by gradually perturbing the initial training samples with Gaussian noise until they are pure white noise (the forward diffusion process), and then training a neural network to predict this perturbation from noisy images in order to reconstruct clean samples (the reverse process; see also Fig.~\ref{fig:schematic}a).
As an extension, including additional input fields during the reverse process allows the model to learn a conditional probability distribution $p_\mathrm{data}(\mathbf{x} \, | \, c)$ given the inputs $c$.
Such models have proved highly successful at image generation tasks \citep[e.g.,][]{ho_denoising_2020}, but they are also starting to be applied to problems from climate science and oceanography such as statistical downscaling \citep[e.g.,][]{ling_diffusion_2024,lopez-gomez_dynamicalgenerative_2025,mardani_residual_2025,aich_conditional_2026}, probabilistic forecasting \citep[e.g.,][]{yoon_deterministic_2023,li_generative_2024,price_gencast_2024,shi_codicast_2025}, and emulation of numerical models \citep[e.g.,][]{bassetti_diffesm_2024,meunier_learning_2025,bouabid_scorebased_2026}.

Whereas diffusion models have been previously employed to emulate single numerical weather prediction and climate models, their application to representing climate variable distributions simulated by multi-model ESM ensembles, particularly for ocean biogeochemical variables such as oxygen, has been less explored.
In this study, we train and evaluate conditional diffusion models on simulation outputs from the CMIP6 ensemble.
Our aim is to represent the conditional distribution $p(\mathrm{O}_2 \, | \, T, \, S, \, \ldots)$ of ocean oxygen concentrations, $\mathrm{O}_2$, given physical input tracers, such as temperature $T$ and salinity $S$, across the model ensemble.

The remainder of this paper is structured as follows: the data and model architecture are detailed in Section \ref{sec:methods};
followed by a test of the sensitivity to the choice of predictor variables (Section \ref{sec:predictors}), an evaluation of seasonal predictions (Section \ref{sec:seasonal}) and of out-of-sample prediction skill (Section \ref{sec:ood}), a validation using observational data (Section \ref{sec:obs_validation}), a comparison to simpler baseline models (Section \ref{sec:baselines_results}), and finally an extrapolation of oxygen predictions to a larger ensemble size (Section \ref{sec:extrapolation}).
Conclusions are discussed in Section \ref{sec:discussion}.

\section{Data and methods}\label{sec:methods}

\subsection{Climate model outputs and observation-based datasets}

\subsubsection{CMIP6 climate model outputs}

We use outputs from 51 ESMs participating in CMIP6, of which 18 provide oxygen outputs to be used as a training dataset.
We used outputs from idealized CO$_2$ forcing scenarios: the \texttt{1pctCO2} scenario with atmospheric CO$_2$ gradually increasing from its preindustrial value by 1\% per year, and the \texttt{abrupt-4xCO2} scenario with an abrupt quadrupling of atmospheric CO$_2$ from its preindustrial value.
The variables used are ocean potential temperature (\texttt{thetao}), salinity (\texttt{so}), oxygen concentration (\texttt{o2}), dissolved inorganic carbon concentration (\texttt{dissic}), and zonal and meridional surface wind stress (\texttt{tauuo} and \texttt{tauvo}).
The number of ESMs and the number of ensemble members per ESM vary across forcing scenarios and variable requirements due to differences in data availability in the CMIP6 archive.
In general, we used as many ensemble members as were available for a given combination of variables and forcing scenario.
The number of ensemble members available from each model is detailed in Table \ref{tab:cmip6_models}.

\subsubsection{Observation-based datasets}

We use observation-based gridded products of ocean temperature, salinity, oxygen concentration, and wind stress from several sources.
These datasets are detailed in Table \ref{tab:obs_datasets}.
We processed all observation-based fields in the same manner as the model output fields as described in the next subsection.
For consistency, we calculated mixed layer depth (MLD) from the monthly ECCO $T$ and $S$ fields, as described above, rather than using the direct MLD output from ECCO. 

\begin{table*}[t]
\caption{
Observation-based datasets used in this study.
}\label{tab:obs_datasets}
\begin{center}
\begin{tabular}{l|l|l|l}
\hline\hline
Dataset & Resolution & Variables used & Citation \\
\hline
World Ocean Atlas 2023 & decadal & $T$, $S$, O$_2$ & \cite{garcia_world_2024} \\
GOBAI-O2 & annual & $T$, $S$, O$_2$ & \cite{sharp_gobaio2_2023} \\
ECCO state estimate & annual & $T$, $S$, MLD & \cite{eccoconsortium_ecco_2024} \\
ERA5 reanalysis & annual & $\tau_{x,y}$ & \cite{hersbach_era5_2020} \\
\hline
\end{tabular}
\end{center}
\end{table*}

\subsubsection{Data preprocessing}

We remapped all model- and observation-based fields to a common 1-degree latitude-longitude grid using bilinear interpolation, yielding a resolution of $(H, W) = (180, 360)$.
To assess the sensitivity of our results to seasonal processes, we consider both monthly mean fields (the native time resolution of the model outputs) as well as annually averaged fields.
In addition, we consider both depth-integrated fields as well as fields on multiple isopycnal layers, as described below.
This processing yields four possible combinations of time resolution (monthly vs. annual mean) and depth resolution (depth-integrated vs. isopycnal layers).
The main analysis in this study focuses on the case of annual mean depth-integrated fields, but we discuss results for all four combinations, with corresponding figures shown in the Appendix. 

We obtained depth-integrated fields by integrating the variables over the top 2000m of the ocean to capture the main pycnocline and the locations of the major OMZs.
Where provided, we used native grid cell bounds to compute the correct cell thickness as weights for the vertical integration, otherwise we inferred the bounds from the depth levels themselves.

For the isopycnal case, we remapped the variables from depth to density coordinates as follows: i) we computed potential density $\sigma_1$ from potential temperature and salinity using the TEOS-10 software toolbox \citep{mcdougallGettingStartedTEOS102011}; ii) for each model, ten target density levels were defined as volume-weighted quantiles (2nd--98th percentile) of that model's density climatology in the first 100 years of its piControl run; iii) tracers were conservatively remapped from depth onto the target density surfaces.
The use of density quantiles instead of fixed density values allows the faithful intercomparison of models that may differ in their climatological stratification and overall density bias.

We calculated monthly mixed layer depth (MLD) as the minimum depth where the local potential density $\sigma_0$ exceeds its surface value by \SI{0.03}{\kilo\gram\per\cubic\meter} \citep{deboyermontegut_mixed_2004},
with potential density obtained from monthly temperature and salinity fields using TEOS-10 \citep{mcdougallGettingStartedTEOS102011}.
In the annual case, we then calculated annual maximum MLD (MLD$_\mathrm{max}$) by taking the annual maximum value of the monthly mean MLD field to quantify the deep winter MLD.

\begin{figure*}[t]
    \centering
    \includegraphics[width=0.95\textwidth]{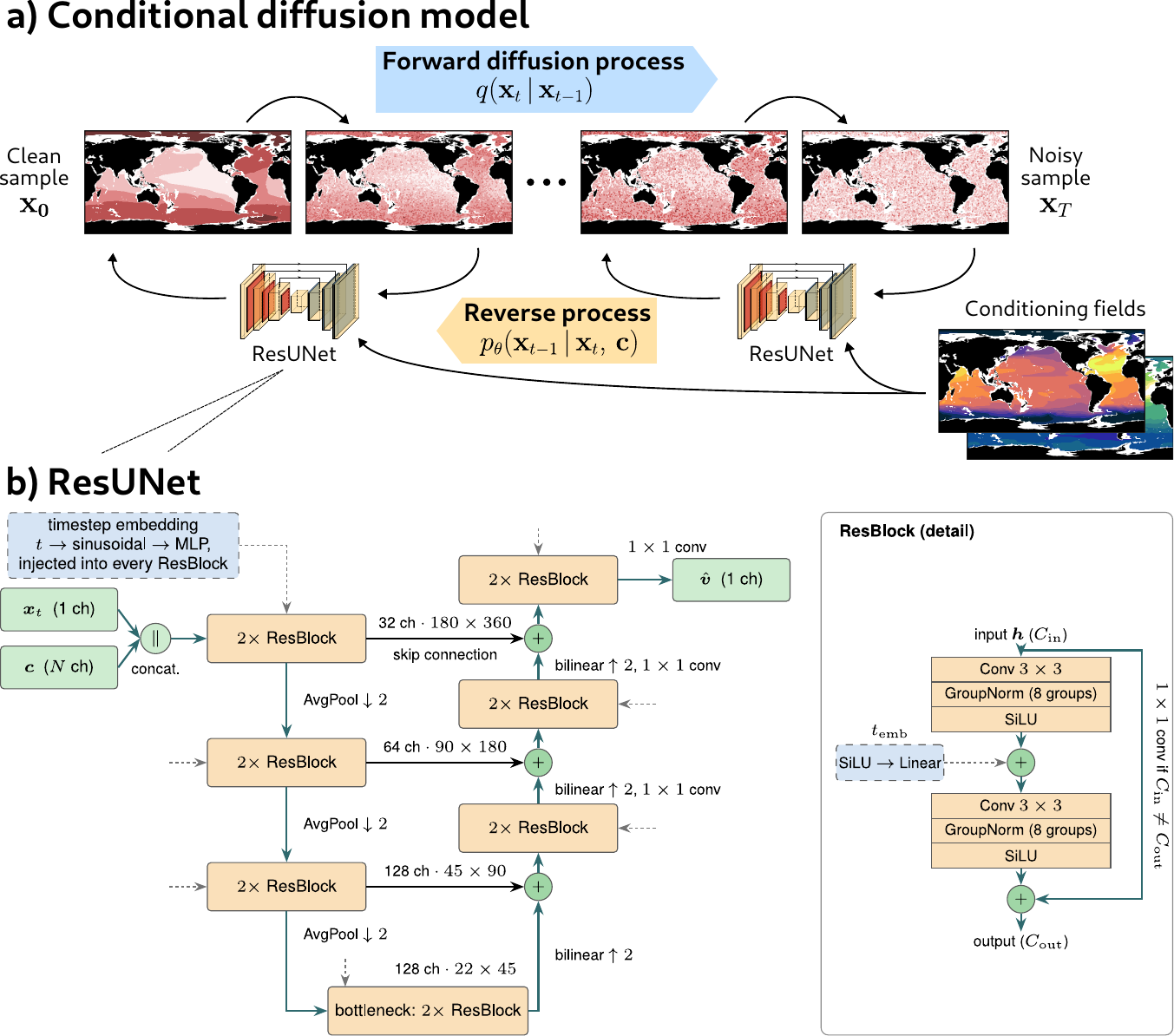}\\
    \caption{
    Schematic of the conditional diffusion model used in this work.
    Panel \textbf{a)}: in the forward process, Gaussian noise is gradually added to the initial map of oxygen concentration $\mathbf{x_0}$ until the sample is pure noise.
    In the reverse process, a U-Net with residual blocks (ResUNet) is then trained to predict the latent velocity variable $\mathbf{v}$ (defined in Eq.~\ref{eq:velocity_parameterization}) given the conditioning fields $\mathbf{c}$, gradually reconstructing a clean sample.
    Panel \textbf{b)} shows a detailed schematic of the ResUNet architecture used to parameterize the reverse process.
    Detailed information for each layer is also given in Table \ref{tab:unet_layers}.
    }\label{fig:schematic}
\end{figure*}

\subsection{Conditional diffusion model}

We use a conditional version of the standard Denoising Diffusion Probabilistic Model \citep[DDPM;][]{ho_denoising_2020} to generate spatial oxygen fields given a set of predictor variables.
As described above, all conditioning and target samples are global tracer maps at 1-degree horizontal resolution, obtained from climate model output.
The diffusion model learns the conditional distribution
\begin{equation}
p(\mathbf{x}_0 \mid \mathbf{c}),
\end{equation}
where $\mathbf{x}_0 \in \mathbb{R}^{H \times W}$ denotes the target oxygen field and $\mathbf{c} \in \mathbb{R}^{C \times H \times W}$ denotes a multi-channel conditioning field.
For the baseline case of depth-integrated $T$ and $S$ as the only predictors, there are $C=2$ channels; when using predictor fields on isopycnal layers this number is multiplied by 10.

The forward diffusion process gradually perturbs $\mathbf{x}_0$ with Gaussian noise:
\begin{equation}
\mathbf{x}_t = \sqrt{\bar{\alpha}_t}\,\mathbf{x}_0 + \sqrt{1 - \bar{\alpha}_t}\,\boldsymbol{\epsilon}, 
\quad \boldsymbol{\epsilon} \sim \mathcal{N}(0, \mathbf{I}),
\end{equation}
where $t \in \{1,\dots,T\}$ is the diffusion timestep and $\alpha_t$ defines the noise schedule as detailed below; we choose $T=1000$ which is a common standard choice in practice \citep{ho_denoising_2020,nichol_improved_2021}.

\subsubsection{Noise schedule}

We employ a cosine noise schedule \citep{nichol_improved_2021}, defined through the cumulative product of noise scaling factors:
\begin{equation}
\bar{\alpha}_t = 
\frac{\cos^2\left( \frac{t/T + s}{1 + s} \cdot \frac{\pi}{2} \right)}
{\cos^2\left( \frac{s}{1 + s} \cdot \frac{\pi}{2} \right)},
\end{equation}
with offset parameter $s = 0.008$. The variance schedule $\{\beta_t\}_{t=1}^T$ is then obtained from
\begin{equation}
\beta_t = 1 - \frac{\bar{\alpha}_t}{\bar{\alpha}_{t-1}},
\end{equation}
with $\alpha_t = 1 - \beta_t$ and $\bar{\alpha}_t = \prod_{i=1}^t \alpha_i$.

A cosine noise schedule has been shown to give improved sample quality over the standard linear schedule \citep{nichol_improved_2021}.

\subsubsection{Model parameterization}

The model is trained to predict the velocity parameterization $\mathbf{v}$:
\begin{equation}
\mathbf{v} = \sqrt{\bar{\alpha}_t}\,\boldsymbol{\epsilon} - \sqrt{1 - \bar{\alpha}_t}\,\mathbf{x}_0,
\label{eq:velocity_parameterization}
\end{equation}
which allows reconstruction of both the denoised target sample $\mathbf{x}_0$ as well as the noise term $\boldsymbol{\epsilon}$ via
\begin{align}
\boldsymbol{\epsilon} &= \sqrt{1 - \bar{\alpha}_t}\,\mathbf{x}_t + \sqrt{\bar{\alpha}_t}\,\mathbf{v}, \\
\mathbf{x}_0 &= \sqrt{\bar{\alpha}_t}\,\mathbf{x}_t - \sqrt{1 - \bar{\alpha}_t}\,\mathbf{v}.
\end{align}

The loss function is the mean squared error between the predicted and true velocity:
\begin{equation}
\mathcal{L} = \mathbb{E}_{\mathbf{x}_0, \mathbf{c}, t, \boldsymbol{\epsilon}} 
\left[ \left\| \mathbf{v}_\theta(\mathbf{x}_t, \mathbf{c}, t) - \mathbf{v} \right\|^2 \right],
\end{equation}
evaluated only over ocean grid cells.

\subsubsection{Network architecture}

The diffusion model is parameterized by a conditional U-Net operating on two-dimensional spatial fields.
The network takes as input the concatenation of the noisy target field $\mathbf{x}_t$ and the (unperturbed) conditioning variables $\mathbf{c}$, and predicts a single output channel corresponding to $\mathbf{v}$.

The architecture is shown schematically in Fig.~\ref{fig:schematic}.
It takes the form of a U-Net \citep{ronneberger_unet_2015} with skip connections between three encoder and decoder layers of width (32, 64, 128), respectively.
Each layer contains two residual blocks with group normalization and SiLU activation.
Downsampling is performed via average pooling and upsampling via bilinear interpolation.
Skip connections are implemented as element-wise addition.
Diffusion time-step information is incorporated via a sinusoidal timestep embedding followed by a multi-layer perceptron, which is injected into each residual block as a channel-wise bias.

\subsubsection{Optimization and training setup}

Training is performed using the Adam optimizer with a learning rate of $10^{-4}$ and a batch size of 16.
Models are trained for up to 1000 epochs with mixed-precision training.

An exponential moving average (EMA) of model parameters is maintained during training with a decay factor of 0.9999.
The EMA weights are used for sampling and evaluation.

\subsubsection{DDIM sampling}

At inference time, samples are generated using a fast sampling procedure from Denoising Diffusion Implicit Models \citep[DDIM;][]{song_denoising_2022}.
Starting from Gaussian noise $\mathbf{x}_T \sim \mathcal{N}(0, \mathbf{I})$, the model iteratively updates the sample using a non-Markovian stochastic mapping.

Given a sequence of timesteps $\{t_k\}_{k=1}^S$ with $t_1 = T$ and $t_S = 0$, the DDIM update from $t$ to $t'$ is
\begin{equation}
\mathbf{x}_{t'} = \sqrt{\bar{\alpha}_{t'}}\,\mathbf{x}_0 
+ \sqrt{1 - \bar{\alpha}_{t'} - \sigma_t^2}\,\boldsymbol{\epsilon}
+ \sigma_t \mathbf{z},
\end{equation}
where $\mathbf{z} \sim \mathcal{N}(0, \mathbf{I})$ and
\begin{equation}
\sigma_t = \eta \sqrt{\frac{1 - \bar{\alpha}_{t'}}{1 - \bar{\alpha}_t}} 
\sqrt{1 - \frac{\bar{\alpha}_t}{\bar{\alpha}_{t'}}}.
\end{equation}

Here, $\mathbf{x}_0$ and $\boldsymbol{\epsilon}$ are reconstructed from the model prediction as described above, and $\eta$ controls the level of stochasticity.
We use $S=100$ sampling steps and $\eta=0.1$ to achieve a favorable trade-off between sampling speed and sample quality, but the results are not sensitive to these specific values (Fig.~\ref{appfig:sampler_sweep}).

The land--sea mask is enforced at each sampling step to ensure physically consistent outputs over ocean regions.

\subsubsection{Uncertainty calibration}

The spread across oxygen samples generated by the diffusion model, conditioned on the same input fields, reflects uncertainty arising from different internal variability states, climate models, and forcing scenarios in the training set.
In the out-of-sample case, where samples are generated, e.g., for climate models not seen during training, the samples may be underdispersed, i.e., the spread across samples may underestimate the true out-of-sample uncertainty.
To correct this, we apply a spatially varying variance inflation calibration.
For each grid point $(\phi,\lambda)$, we define a calibration factor
\begin{equation}
  \alpha(\phi,\lambda) = \sqrt{\frac{\frac{1}{M}\sum_{m=1}^{M}
    \bigl(y_m(\phi,\lambda) - \bar{x}_m(\phi,\lambda)\bigr)^2}
    {\frac{1}{M}\sum_{m=1}^{M} s_m^2(\phi,\lambda)}},
\end{equation}
where $y_m$ is the ground-truth oxygen field for model $m$, $\bar{x}_m = \frac{1}{N}\sum_{n=1}^{N} x_m^{(n)}$ is the sample mean over the $N$ generated samples, and $s_m^2$ is the corresponding sample variance.
The summation runs over all $M$ climate models in the leave-one-model-out ensemble.
Because $M$ is small, the pointwise estimates of $\alpha$ are noisy; we therefore spatially smooth $\alpha^2$ using a Gaussian kernel (applied in variance space to preserve linearity), with a normalization correction to prevent land contamination at coastlines.
Calibrated samples are then obtained as $\tilde{x}_m^{(n)} = \bar{x}_m + \alpha \cdot (x_m^{(n)} - \bar{x}_m)$.

To verify that the inflation factor generalizes, we fit and validate it in a nested leave-one-model-out loop: for each held-out model, $\alpha(\phi, \lambda)$ is fitted from the errors and spreads of the other $M-1$ models only, and the inflated 50-member ensemble is then scored against the held-out model (years 81--100).
The fit calls for a large correction, with a median $\alpha$ of about 12, confirming that the raw ensemble is too narrow.
Inflation leaves the ensemble mean prediction error unchanged, but improves the probabilistic scores: CRPS improves for 11 of the 14 held-out models (median -18\%), and coverage of the 96\% confidence interval rises from 21\% to 87\%.
Note that the correction acts on the ensemble average: it brings the mean spread-to-error ratio to 1.0 while individual models still span 0.2--2.7, so the calibrated spread should not be read as a per-model uncertainty. 

\subsection{Baselines}
\label{sec:baselines_methods}

We assess the added benefit of the diffusion model over simpler approaches by training four baseline methods of decreasing complexity and evaluating them on the same protocol as the diffusion model: (i) a U-Net trained with a mean squared error (MSE) loss; (ii) a pointwise quantile regression forest (QRF); (iii) a pointwise linear regression; and (iv) a per-cell climatology, the intercept-only submodel of the linear regression model.

The U-Net baseline is built on the same ResUNet backbone as the diffusion model and represents a deterministic variant of the probabilistic generative model.
The pointwise QRF is probabilistic in nature and thus provides a baseline against which to compare the diffusion model's generated spread and uncertainty calibration.
Finally, the linear baseline represents the most naive prediction that still uses the predictors, while its climatological submodel, which uses no predictors at all, provides the floor against which to compare the error metrics from all other methods.

The baselines are trained and evaluated on the main configuration with
annual mean depth-integrated $T$ and $S$ from 1pctCO2 as predictors, both in the in-sample case and the out-of-sample \enquote{leave-one-model-out} (LOMO) case (see Section \ref{sec:baselines_results}).

\subsubsection{Deterministic U-Net}

The U-Net baseline is identical to the ResUNet backbone of the diffusion model, except with one fewer input channel due to the absence of the noisy input channel.
The loss function is the global ocean MSE, and all optimization settings are identical to those of the diffusion model.
The epoch checkpoint is selected by minimizing the validation loss over held-out years to ensure that the baseline is not under-tuned compared to the diffusion model.

\subsubsection{Quantile regression forest}

The QRF baseline \citep{meinshausen_quantile_2006} is an extension of the random forest \citep{breiman_random_2001} which yields an estimation of the full conditional distribution instead of a mean estimate.
It is trained in a pointwise fashion where the training samples consists of each grid cell's predictor values together with a sine/cosine encoding of its latitude and longitude.
A single global forest is fitted to samples subsampled from every training map, so its only spatial dependence is through the location features.
The minimum number of training samples per leaf is selected on the same validation split as the U-Net, using the mean pinball loss \citep{gneiting_strictly_2007}.
For prediction, the forest returns an estimated conditional distribution at every ocean cell, evaluated on $101$ equally spaced percentile levels.
Samples are then generated by inverse transform sampling from these distributions.

\subsubsection{Pointwise linear regression and climatology}

The linear baseline is an pointwise ordinary least squares regression of the target on the predictors at each grid cell, with independent parameters across different grid cells.
Setting its predictor coefficients to zero yields hte even simpler intercept-only climatology submodel, a per-cell climatology that uses no predictors at all.

\begin{figure*}[t]
    \centering
    \includegraphics[width=\textwidth]{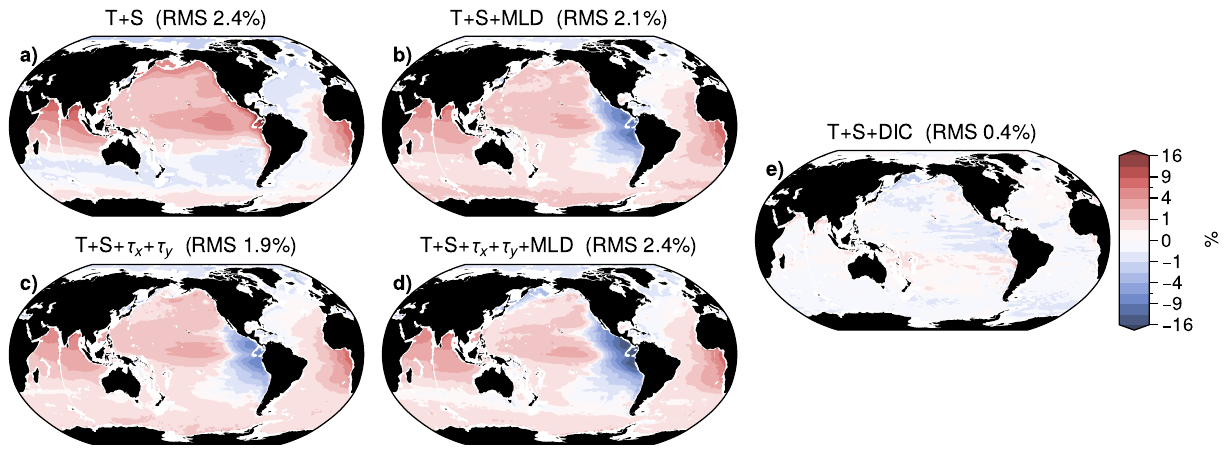}\\
    \caption{
    Sensitivity of local generative skill to the choice of predictor variables (for annual mean depth-integrated fields).
    The maps show the signed relative error across the climate model ensemble averaged over the held-out years 81-100 of the 1pctCO2 scenario.
    The numbers in parentheses state the global root mean squared (RMS) error for each panel.
    \textbf{a)--d)}: physical predictors (0--2000m depth-integrated temperature $T$ and salinity $S$, annual maximum mixed layer depth $\mathrm{MLD}_\mathrm{max}$, as well as zonal and meridional surface wind stress $\tau_x$ and $\tau_y$).
    \textbf{e)}: same as a) but including 0--2000m depth-integrated dissolved inorganic carbon concentration (DIC) as an additional predictor.
    Analogous figures for monthly mean fields and fields on isopycnal layers are provided in the appendix (Fig.~\ref{appfig:predictor_set}).
    }\label{fig:predictor_set}
\end{figure*}

\begin{figure*}[t]
    \centering
    \includegraphics[width=0.8\textwidth]{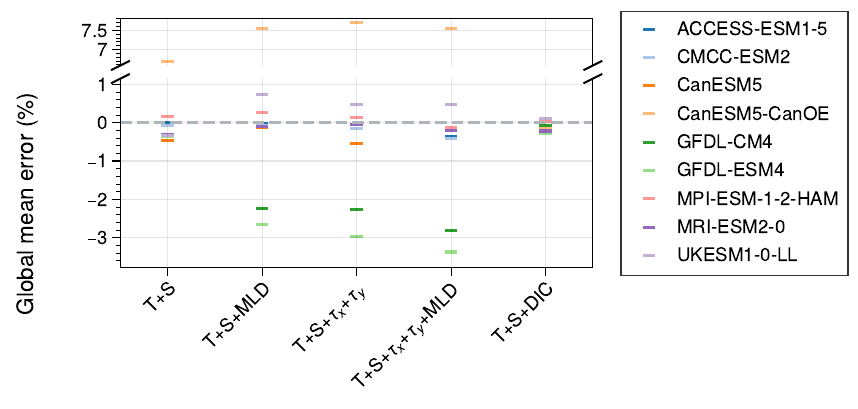}\\
    \caption{
    Sensitivity of global mean generative skill to the choice of predictor variables (for annual mean depth-integrated fields).
    The data and variable combinations are as in Fig.~\ref{fig:predictor_set}, but the colored lines show the area-weighted global mean relative oxygen prediction error for each climate model separately.
    The y-axis is cut to enhance the detail for the group of models with errors close to 0\%.
    Note that, in the right-most column, the outlier model (CanESM5-CanOE) is located in the lower part of the y-axis close to 0\%.
    Analogous figures for monthly mean fields and fields on isopycnal layers are provided in the appendix (Fig.~\ref{appfig:predictor_boxplot}).
    }\label{fig:predictor_boxplot}
\end{figure*}

\section{Sensitivity of diffusion model skill to choice of predictor variables}\label{sec:predictors}

As a first step, we evaluate the sensitivity of the diffusion model's skill to the choice of predictor variables.
Figure \ref{fig:predictor_set} shows the mean signed relative error of oxygen predictions across the climate model ensemble for different choices of the predictor variables, using annual mean depth-integrated fields.
For each choice of predictors (i.e., for each panel in Fig.~\ref{fig:predictor_set}), we trained a diffusion model on CMIP6 outputs from years 1--80 and 100--150 of the 1pctCO2 scenario.
Outputs from all available climate model members were used both in training and evaluation.
Then, for each climate model, we computed the percentage error of the predicted oxygen map, using predictors from years 81--100 of the 1pctCO2 scenario which were held out from the training set, relative to the ground-truth oxygen field in those years.
Averaging these errors over the model ensemble and over the 20 held-out years yields the error maps shown in Fig. ~\ref{fig:predictor_set}.

Panels a)--d) of Fig.~\ref{fig:predictor_set} show the signed relative error using different combinations of physical variables as predictors.
Using only depth-integrated temperature and salinity as predictors (Fig.~\ref{fig:predictor_set}a) gives a global RMS of 2.4\%, and the error is highest in the tropical OMZs and in the subpolar North Pacific, with values over 10\% in these regions, while the errors are much smaller across the Atlantic and Southern Oceans.
Including additional physical predictors, such as annual maximum mixed-layer depth (MLD$_\mathrm{max}$, Fig.~\ref{fig:predictor_set}b) and surface wind stress ($\tau_x$ and $\tau_y$, Fig.~\ref{fig:predictor_set}c), or both (Fig.~\ref{fig:predictor_set}d), yields only marginal improvements in overall skill, although the sign of the Pacific OMZ error flips from positive to negative when using more physical predictor variables than just $T$ and $S$.

Panel \ref{fig:predictor_set}e) shows the signed relative error using $T$ and $S$ as in panels a)--d), but also including depth-integrated dissolved inorganic carbon (DIC) as an additional biogeochemical predictor.
Adding DIC as a predictor strongly reduces the relative error in all regions, with a global RMS of only 0.4\%.

Performing the same analysis with monthly mean fields as well as fields on isopycnal layers yields similar results (Fig.~\ref{appfig:predictor_set}).
Compared to the annual mean depth-integrated setting of Fig.~\ref{fig:predictor_set}, the prediction error is slightly decreased when using monthly mean data, both depth-integrated and on isopycnal layers.
In contrast, using annual mean data on isopycnal layers leads to degraded skill (Fig.~\ref{appfig:predictor_set}f--j).

Three main conclusions can be reached from these observations: i) the inclusion of DIC as an additional predictor in the diffusion model leads to a substantial reduction in error regardless of temporal and vertical resolution; ii) the inclusion of further physical variables beyond temperature and salinity (i.e., mixed layer depth or surface wind stress) only yields minor decreases in error; and iii) models trained on monthly mean data show better skill than models trained on annual mean data, while the use of isopycnal layers is only effective at monthly resolution.

The same sensitivity test is shown in terms of global mean percentage error in Fig.~\ref{fig:predictor_boxplot}.
The diffusion model is trained on full two-dimensional conditioning and target fields, and the ResUNet's loss function does not explicitly minimize the global mean oxygen concentration error, but a skillful prediction should nevertheless yield a global mean as close as possible to the ground truth.  

Figure \ref{fig:predictor_boxplot} reveals a feature of the predicted oxygen fields not seen in Fig.~\ref{fig:predictor_set}: for the physical predictor variables, the predictions from the model CanESM5-CanOE are outliers, with significantly higher relative global mean errors than those from the other climate models.
This behavior disappears when DIC is added as an additional predictor, with errors for CanESM5-CanOE falling within the range of the other models' errors. 
The inclusion of additional physical predictors (MLD, $\tau_{x,y}$) leads to a degradation of predictive skill for some models, while the inclusion of DIC strongly reduces errors for all models.

The outlier behavior of CanESM5-CanOE persists when using monthly mean and vertically resolved data (Fig.~\ref{appfig:predictor_boxplot}).
The use of data on isopycnal layers generally increases the spread in global mean error across models, and for most predictor variable combinations there is a partial compensation in errors between the outlier model and the group of non-outlier models.

Nevertheless, the conclusions regarding the choice of predictors remain the same: including DIC reduces prediction errors, but including further physical variables or depth information does not substantially improve the skill. 
For these reasons, we will focus on depth-integrated temperature and salinity as predictors for the remainder of this paper.
Dissolved inorganic carbon, although a highly skillful predictor, is only available for models with a biogeochemistry component (the same models that simulate O$_2$; see Table \ref{tab:cmip6_models}), and would thus not be available for the extrapolation to climate models with only physical output variables (Section \ref{sec:extrapolation}).

\begin{figure*}[t]
    \centering
    \includegraphics[width=\textwidth]{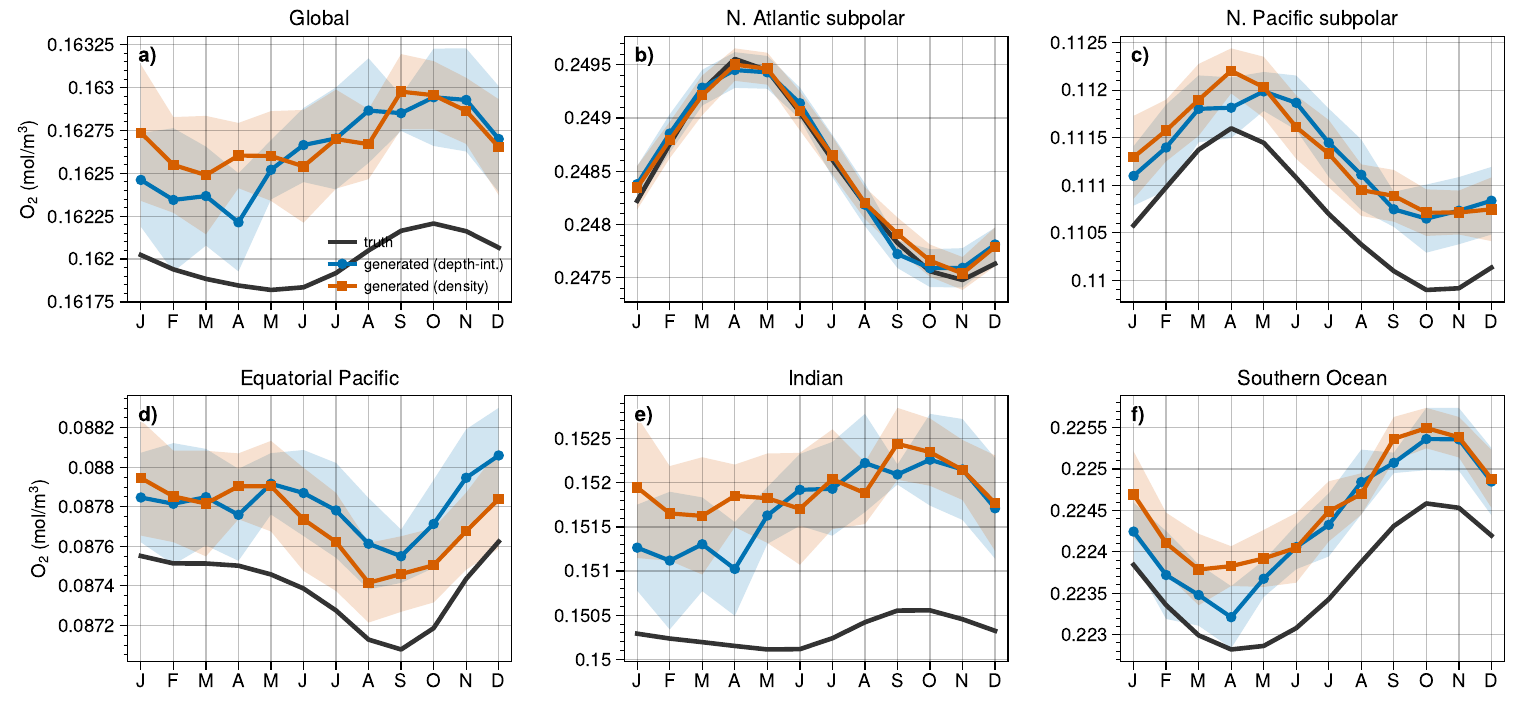}\\
    \caption{
    Seasonal cycle of true and generated regional oxygen fields using monthly data.
    Each panel includes the multi-model mean ground truth seasonal cycle (black), and the multi-model mean generated seasonal cycle using diffusion models trained on depth-integrated monthly data (blue) and monthly data on isopycnal layers (orange).
    The shading shows one standard deviation across the ensemble.
    An analogous figure with the annual mean bias subtracted in each panel is provided in the appendix (Fig.~\ref{appfig:seasonal_cycle_anomaly}).
    }\label{fig:seasonal_cycle_absolute}
\end{figure*}

\section{Seasonal generation with monthly data}\label{sec:seasonal}

Using monthly mean data for the predictor and target fields, we can evaluate the fidelity of the reconstructed seasonal cycle in oxygen concentration.
Figure \ref{fig:seasonal_cycle_absolute} shows ground truth and generated oxygen seasonal cycles averaged over the global ocean (Fig.~\ref{fig:seasonal_cycle_absolute}a) and over selected regions (Fig.~\ref{fig:seasonal_cycle_absolute}b--f).
An analogous figure with the annual mean bias subtracted in each panel (to focus the comparison between generated and ground truth fields on the phase and amplitude of the seasonal cycle) is provided in the appendix (Fig.~\ref{appfig:seasonal_cycle_anomaly}).
In general, both the phase and amplitude of the regional oxygen seasonal cycle is well represented by the generated fields.
The lag between predicted and ground truth seasonal cycles does not exceed 1 month in any region, and the seasonal amplitude ratio between generated and ground truth fields is indistinguishable from 1 in 9 out of the 12 cases in Fig.~\ref{fig:seasonal_cycle_absolute}.
The most exact agreement is found in the North Atlantic subpolar region, which is also the region with the smallest annual mean bias.
The seasonal skill is similar for both choices of vertical resolution (i.e., depth-integrated predictor fields in blue vs. fields on isopycnal layers in orange).

\section{Out-of-sample prediction capabilities}\label{sec:ood}

An important concern for the application of machine learning methods to climate data is their generalization to states not seen in the training data, such as perturbed future climate states \citep[e.g.,][]{beucler_climateinvariant_2024}. 
In this subsection, we evaluate the skill of the diffusion model at extrapolating to such out-of-sample states in the context of unseen forcing scenarios (i.e., training on model outputs from one scenario and evaluating predictions on another), and of unseen climate models (i.e., training on outputs from a subset of the climate model ensemble, and evaluating on the remaining models).
The latter test is particularly challenging and specific to our setup, as we train our diffusion models on outputs from an ensemble of different climate models instead of on a single data source such as a reanalysis dataset.

\begin{figure*}[t]
    \centering
    \includegraphics[width=0.75\textwidth]{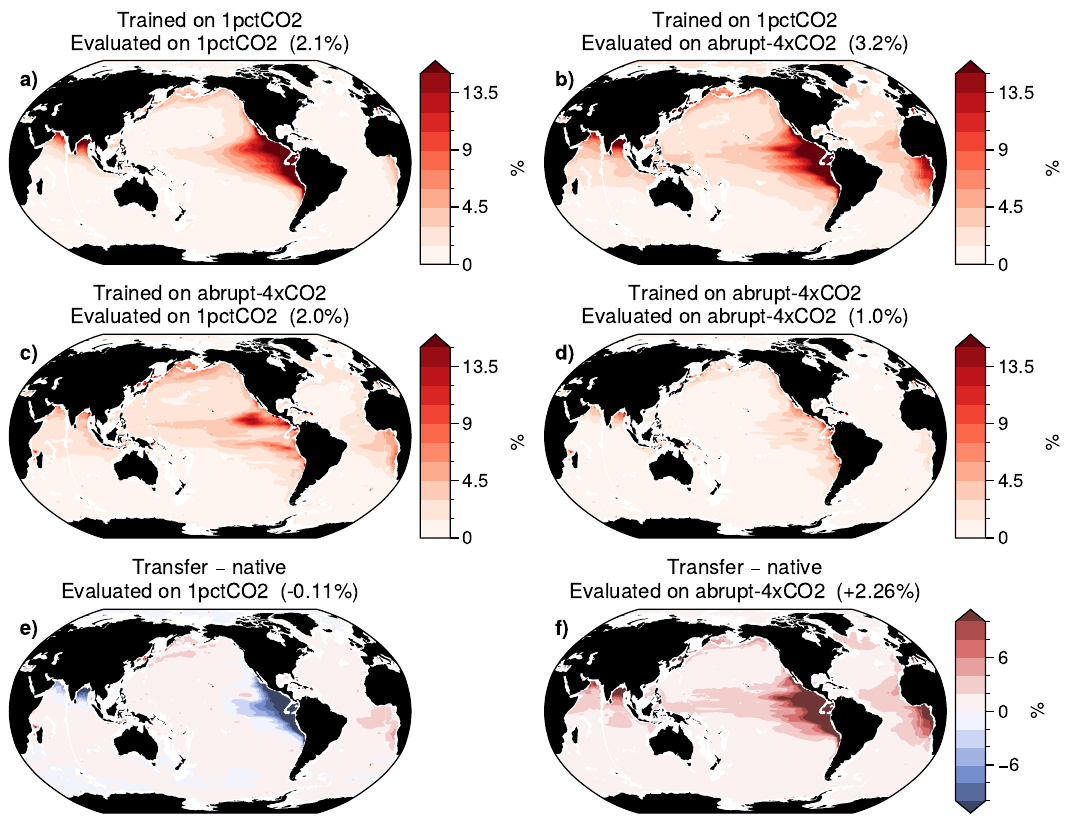}\\
    \caption{
    Generative skill in applications to out-of-sample forcing scenarios (for annual mean depth-integrated fields).
    Panels a)--d) show the mean absolute percentage error (MAPE) across the climate model ensemble in the held-out years 81--100, for a diffusion model trained and evaluated on outputs for specific scenarios (1pctCO2 and abrupt-4xCO2).
    \textbf{a)}, trained and evaluated on 1pctCO2 (in-sample).
    \textbf{b)}, trained on 1pctCO2 and evaluated on abrupt-4xCO2 (out-of-sample).
    \textbf{c)}, trained on abrupt-4xCO2 and evaluated on 1pctCO2 (out-of-sample).
    \textbf{d)}, trained and evaluated on abrupt-4xCO2 (in-sample).
    \textbf{e)}, difference between panels c) and a).
    \textbf{f)}, difference between panels d) and b).
    The numbers in parentheses state the global mean MAPE.
    Analogous figures for monthly mean fields and fields on isopycnal layers are provided in the appendix (Fig.~\ref{appfig:scenario_transfer}).
    }\label{fig:scenario_transfer}
\end{figure*}

\subsection{Out-of-sample forcing scenarios}

Figure \ref{fig:scenario_transfer} shows the mean absolute percentage error (MAPE) across the climate model ensemble when predicting oxygen fields for both in-sample and out-of-sample forcing scenarios, using only annual mean depth-integrated temperature and salinity as predictors.
The error pattern is similar for all considered combinations of forcing scenarios used for training and evaluation, and resembles the patterns already found in the in-sample case in Fig.~\ref{fig:scenario_transfer}: elevated relative errors in the low-oxygen tropical OMZs, and high skill in the Atlantic and Southern Oceans.
This indicates that the diffusion model can generate reasonable samples even when the conditioning fields are from climate states never seen during training.
Perhaps surprisingly, the out-of-sample skill when training on abrupt-4xCO2 data and evaluating on 1pctCO2 data (Fig.~\ref{fig:scenario_transfer}c) is better than in-sample skill when both training and evaluating on 1pctCO2 data (Fig.~\ref{fig:scenario_transfer}a).
The best skill out of these combinations is achieved for the in-sample abrupt-4xCO2 case (Fig.~\ref{fig:scenario_transfer}d).
In all combinations, the global mean out-of-sample MAPE does not exceed 3.5\%.

Out-of-sample skill is enhanced when using monthly mean fields (Fig.~\ref{appfig:scenario_transfer}a--d) and fields on isopycnal layers (Fig.~\ref{appfig:scenario_transfer}i--l).
As for the in-sample predictions discussed in the previous section, the annual mean depth-integrated case exhibits the worst skill out of the four combinations of temporal and vertical resolution.
There is no single combination that performs better out-of-sample than all others, but monthly data systematically outperforms annual data.

\begin{figure*}[t]
    \centering
    \includegraphics[width=\textwidth]{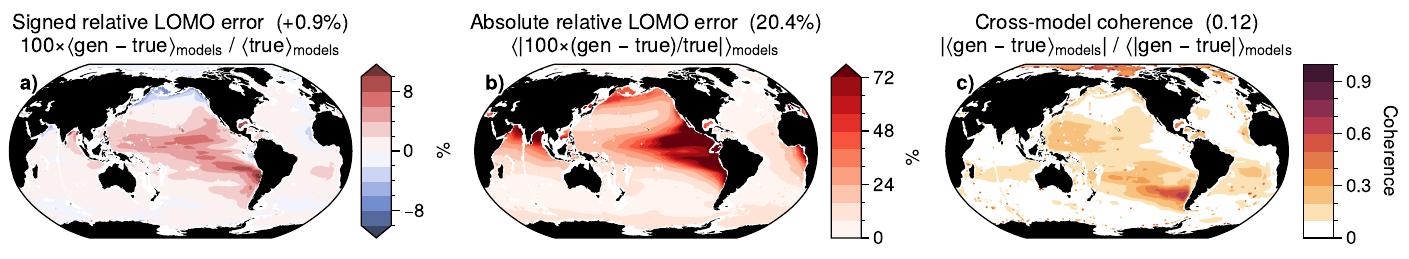}\\
    \caption{
    Generative skill in applications to out-of-sample climate models (for annual mean depth-integrated fields).
    \textbf{a)}, signed relative error across the climate model ensemble of oxygen predictions for each climate model in years 81--100 of the 1pctCO2 scenario, for a diffusion model trained on outputs from all other climate models.
    \textbf{b)}, same as a) but as the mean average percentage error (MAPE).
    \textbf{c)}, coherence of the error's sign across the model ensemble (0 indicates cancellation of positive and negative errors, 1 indicates coherent errors).
    The numbers in parentheses in each panel's title indicate the global mean value.
    Analogous figures for monthly mean fields and fields on isopycnal layers are provided in the appendix (Fig.~\ref{appfig:lomo_error}).
    }\label{fig:lomo_error}
\end{figure*}

\begin{figure*}[t]
    \centering
    \includegraphics[width=0.8\textwidth]{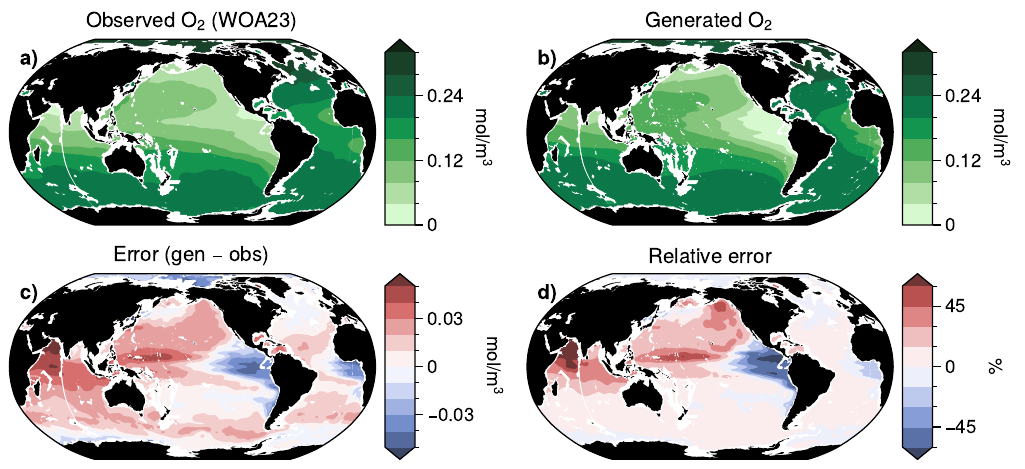}\\
    \caption{
    Observational validation of learned generative conditional distribution (for annual mean depth-integrated fields).
    \textbf{a)}, ground truth oxygen field in years 2004--2017 from the World Ocean Atlas 2023 \citep{garcia_world_2024}.
    \textbf{b)}, generated oxygen field using 0--2000m depth-integrated temperature and salinity from the World Ocean Atlas as predictors.
    \textbf{c)}, oxygen prediction error in physical units (panel b minus a).
    \textbf{d)}, relative oxygen prediction error in percent (panel b minus a, divided by a).
    Analogous figures for monthly mean fields, fields on isopycnal layers, and alternative predictor variable combinations are provided in the appendix (Fig.~\ref{appfig:obs_validation}).
    }\label{fig:obs_validation}
\end{figure*}

\begin{figure*}[t]
    \centering
    \includegraphics[width=\textwidth]{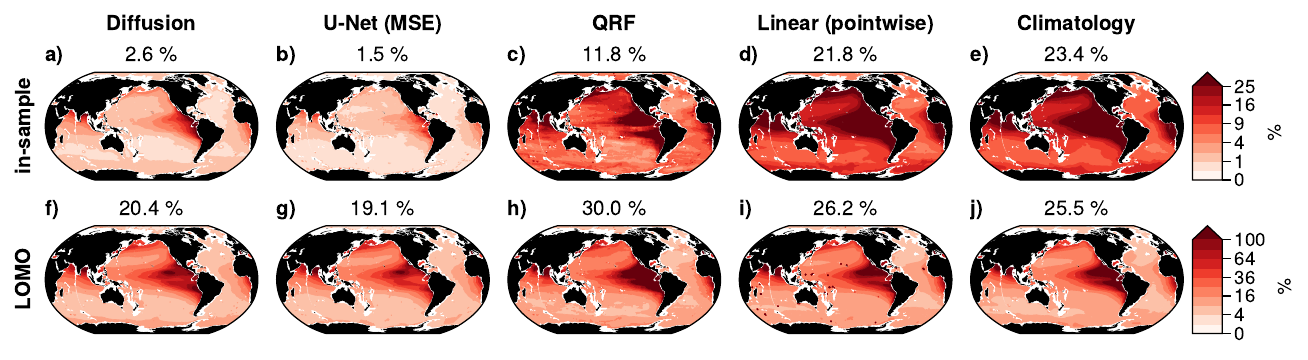}\\
    \caption{
    Comparison of in-sample and leave-one-model-out (LOMO) errors between the diffusion model and baseline methods.
    Each panel shows the mean average percentage error (MAPE) in percent of oxygen predictions by the specified method.
    The number given in each panel title is the global average MAPE.
    The color scale is nonlinear and is shared within each row but differs between the two rows.
    All methods are trained and evaluated using annual mean depth-integrated $T$ and $S$ from the 1pctCO2 scenario as predictors.
    }\label{fig:baselines}
\end{figure*}

\subsection{Out-of-sample climate models}

Figure \ref{fig:lomo_error} shows the diffusion model's prediction skill under extrapolation to out-of-sample climate models not used in training.
These training runs were performed using a \enquote{leave-one-model-out} (LOMO) setup: for each climate model in the ensemble, a diffusion model was trained on the outputs of all other models and evaluated using conditioning fields and ground-truth oxygen from the held-out model.
As before, training used years 1--80 and 100--150 of the 1pctCO2 scenario, and evaluation used the held-out years 81--100 of the same scenario.
We evaluate the resulting generated samples in terms of their average error across the model ensemble, both in terms of mean signed relative error (Fig.~\ref{fig:lomo_error}a) and mean absolute percentage error (Fig.~\ref{fig:lomo_error}b), both relative to each climate model's ground truth, as well as in terms of the coherence of errors for individual models (Fig.~\ref{fig:lomo_error}c). 
The overall pattern of the absolute error (Fig.~\ref{fig:lomo_error}b) is similar to the in-sample case and the out-of-sample forcing scenario, with highest errors in low-oxygen OMZs and lowest in the Atlantic and Southern Oceans.
However, the signed error map (Fig.~\ref{fig:lomo_error}a) shows that the error in the eastern part of the tropical Pacific OMZ is close to zero because of a compensation between models with positive and negative errors.
This is also evident from the coherence map (Fig.~\ref{fig:lomo_error}c), which shows that coherence in this region is small (errors from different models compensate), whereas coherence is larger in the western and southern Pacific (i.e., errors have the same sign).
As expected, the magnitude of these LOMO errors is larger than in the in-sample case, with MAPE values of up to 50\% of the ground truth in regions with the lowest climatological oxygen concentration.
This pattern is similar for diffusion models trained on monthly mean data and data on isopycnal layers (Fig.~\ref{appfig:lomo_error}), but the LOMO error is reduced when using monthly data, similar to the in-sample behavior found above. 

\section{Observational validation of learned generative conditional distribution}\label{sec:obs_validation}

As a way to validate the conditional distribution $p(\mathrm{O}_2 \, | \, T, \, S)$ learned by the diffusion model, we apply the diffusion model to temperature and salinity conditioning fields obtained from observations and compare the generated oxygen field to observations.
Since the diffusion model is trained exclusively on climate model outputs, it is not expected to perfectly capture observations, as any biases in the training data will necessarily be reflected in the trained model.
Instead, knowing that the diffusion model is sufficiently capable of representing relationships between physical predictors and oxygen, this comparison offers a novel way to evaluate the distribution of oxygen as a function of temperature and salinity across the climate model ensemble.

Figure \ref{fig:obs_validation} shows the ground truth oxygen field from the World Ocean Atlas (WOA) 2023 \citep{garcia_world_2024} along with the oxygen field generated by the diffusion model given depth-integrated temperature and salinity, also from the WOA23.
Overall, the generated field accurately captures the global and regional structures of the true oxygen field.
Oxygen is overestimated in the North Pacific and Indian oceans and underestimated in the eastern tropical Pacific OMZ, while errors are much smaller in the Atlantic and Southern Oceans.
This pattern is generally robust across observational datasets and additional predictor variables, as shown in Fig.~\ref{appfig:obs_validation} using $T$, $S$, and MLD from the ECCO state estimate \citep{forget_ecco_2015}, $\tau_{x,y}$ from ERA5 \citep{hersbach_era5_2020}, and DIC from MOBO-DIC \citep{keppler_recent_2023}.

\section{Comparison to baseline methods}\label{sec:baselines_results}

In order to assess the added benefit of the diffusion model over simpler approaches, we compare prediction errors of the diffusion models to the four baseline methods detailed in the Methods section.
Figure \ref{fig:baselines} shows the mean absolute percentage error (MAPE) of the diffusion model and each of the four baseline methods, both in the in-sample and the leave-one-model-out setting.
All methods are trained and evaluated using annual mean depth-integrated $T$ and $S$ from the 1pctCO2 scenario as predictors.

In the in-sample case, the two convolutional architectures (the diffusion model and the U-Net) vastly outperform the pointwise QRF and the basic linear and climatological baselines.
The U-Net has a slightly lower global mean MAPE than the diffusion model (1.5\% for the U-Net vs. 2.6\% for diffusion), while both have much smaller errors than the QRF (11.8\%), linear regression (21.8\%), and climatology (23.4\%).

In the leave-one-model-out case, the diffusion model and U-Net also have lower errors than the other baselines (20.4\% and 19.1\% global mean MAPE, respectively), but the gap to the next-best method is smaller (the climatological baseline with 25.5\%).

Overall, the diffusion model is similar in skill to the deterministic U-Net, and both outperform the pointwise nonlinear and linear baselines.
The fact that the U-Net slightly outperforms the diffusion model is not unexpected, since the U-Net is explicitly trained to minimize global MSE.
The chief benefit of the diffusion model thus lies in its combination of skillful mean predictions (similar to the U-Net) with the ability to produce ensembles of predicted samples given the same conditioning features (similar to the QRF, but with smaller errors).

\begin{figure*}[t]
    \centering
    \includegraphics[width=0.95\textwidth]{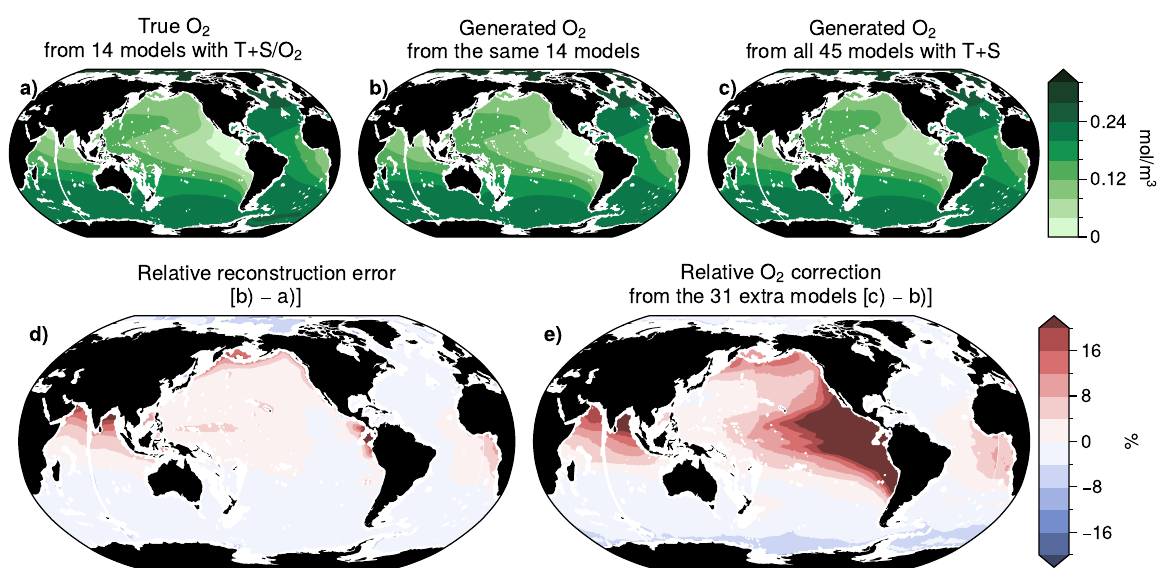}\\
    \caption{
    Oxygen generation based on physical predictors for a larger multi-model ensemble (for annual mean depth-integrated fields).
    \textbf{a)}, ensemble mean ground truth oxygen field in the 14 CMIP6 models with biogeochemistry outputs.
    \textbf{b)}, ensemble mean of generated oxygen fields for each of the 14 models in panel a).
    \textbf{c)}, ensemble mean of generated oxygen fields for a larger set of 45 models that provide at least temperature and salinity outputs.
    \textbf{d)}, relative prediction error for 14 models with biogeochemistry outputs (panel b minus a).
    \textbf{e)}, relative difference in ensemble mean oxygen field predicted for the set of 45 models vs. the field predicted for the set of 14 models (panel c minus b).
    Analogous figures for monthly mean fields and fields on isopycnal layers are provided in the appendix (Fig.~\ref{appfig:extrapolation}).
    }\label{fig:extrapolation}
\end{figure*}

\section{Extrapolation to larger multi-model ensemble}\label{sec:extrapolation}

Having assessed the diffusion model's skill at out-of-sample extrapolation tasks (Section \ref{sec:ood}) and its consistency with observations (Section \ref{sec:obs_validation}), we now turn to the task of extrapolating to a larger multi-model ensemble.
Specifically, we train a diffusion model on all 14 models that have oxygen outputs available (as before), and then generate oxygen fields using conditioning input fields from the larger set of all 45 CMIP6 models with at least temperature and salinity outputs available, including models without any biogeochemical outputs (see Table \ref{tab:cmip6_models}).
The outcome of this extrapolation is shown in Figure \ref{fig:extrapolation}, including the ground truth oxygen field for the smaller set of models with biogeochemistry (panel a), the generated oxygen field for this smaller set (panel b), and the generated oxygen field for the larger set of 45 models (panel c).
As shown previously, the in-sample signed relative error for the smaller set of models is small (panel d), with MAPE below 10\% everywhere.

Comparing the generated ensemble mean oxygen fields between the larger and smaller sets of models, we find that the larger set of models implies a higher oxygen concentration in the eastern tropical Pacific OMZ (Fig.~\ref{fig:extrapolation}e) by up to 50\% of the baseline values.
This suggests that the physical conditions simulated by the larger set of models are consistent with higher oxygen concentrations in OMZs than those simulated by the small set of models with explicit biogeochemical components.
Note, however, that this inference necessarily rests on the distribution $p(\mathrm{O}_2 \, | \, T, \, S)$ learned by the diffusion model by training on the set of 14 models with biogeochemistry outputs.

When training the diffusion model on monthly data, as well as data on isopycnal layers, this correction from the larger set of climate models is qualitatively consistent, with a positive correction in the tropical Pacific OMZ in all cases (Fig.~\ref{appfig:extrapolation}). 
The strength of this correction differs between setups, with the smallest correction for the annual mean isopycnal case which also had the poorest skill out of the four combinations of temporal and vertical resolution.
However, the three best combinations (annual and monthly depth-integrated fields, and monthly fields on isopycnal layers) all agree on a sizable upward regional correction in the eastern tropical Pacific.

\begin{figure*}[t]
    \centering
    \includegraphics[width=0.95\textwidth]{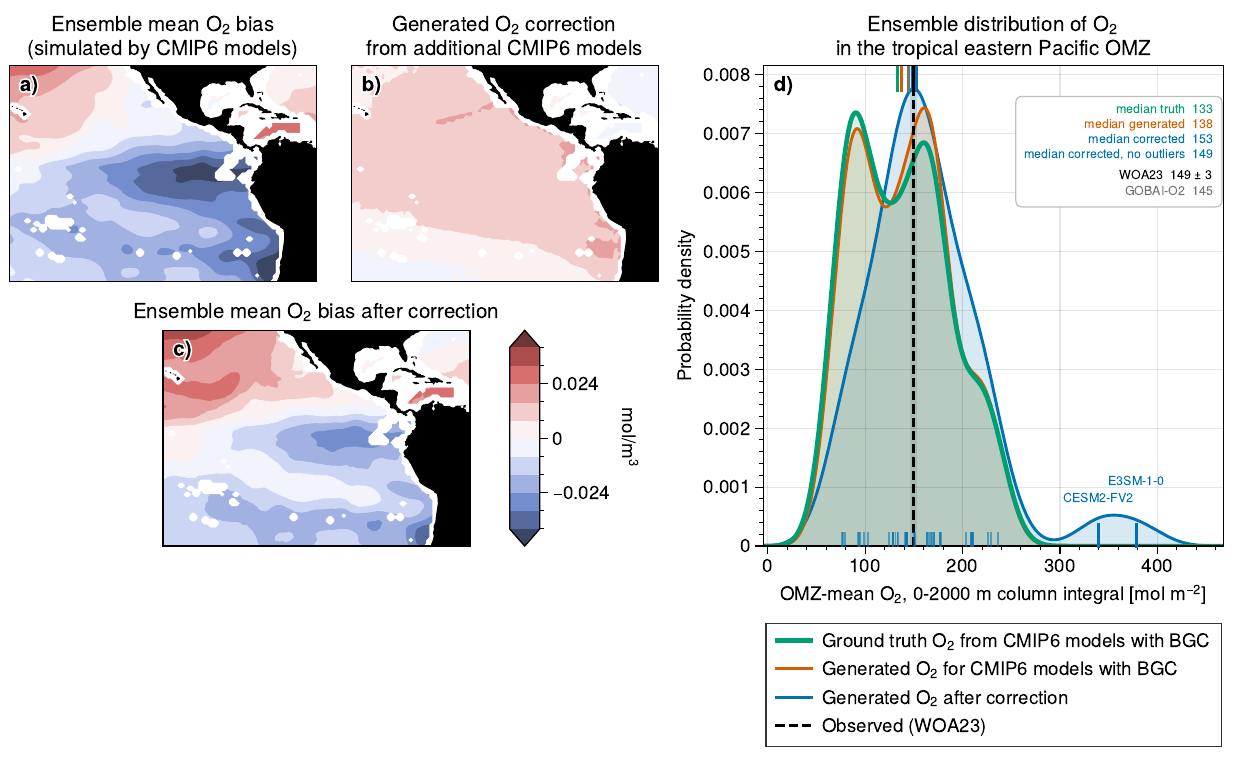}\\
    \caption{
    Implications of oxygen extrapolation to larger ensemble size on the eastern tropical Pacific Oxygen Minimum Zone (for annual mean depth-integrated fields).
    \textbf{a)}, ensemble mean oxygen bias in CMIP6 relative to observations.
    \textbf{b)}, oxygen correction implied by extrapolation to larger ensemble (same as Fig.~\ref{fig:extrapolation}e but in absolute units).
    \textbf{c)}, ensemble mean oxygen bias in CMIP6 relative to observations after adding the correction in panel b).
    \textbf{d)}, inter-model distribution of eastern tropical Pacific oxygen concentration before and after applying the correction.
    The curves are estimated from histograms of individual model values via Gaussian kernel density estimation, with a fixed kernel bandwidth of 0.4 times each curve's own standard deviation.
    The OMZ is defined as the area within 30° of the equator in the Pacific whose observed 0–2000 m column-integrated oxygen is below \SI{200}{\mol\per\square\meter}.
    The dashed vertical line indicates the observational value from the World Ocean Atlas 2023 \citep{garcia_world_2024}, and the WOA23 uncertainty stated in the legend is its standard error averaged over the region assuming fully correlated errors.
    The short blue lines at the bottom of the panel indicate individual model values including two highlighted outlier models (CESM2-FV2 and E3SM-1-0).
    The short colored lines at the top of the panel indicates the median of each distribution.
    Analogous versions of panel d) for monthly mean fields and fields on isopycnal layers are provided in the appendix (Fig~\ref{appfig:extrapolation_omz}).
    }\label{fig:extrapolation_omz}
\end{figure*}

\subsection{Eastern Pacific OMZ bias}

Given the prominence of the eastern tropical Pacific OMZ in the upward oxygen correction identified in Fig.~\ref{fig:extrapolation}e), we now focus on this region more directly.
Figure \ref{fig:extrapolation_omz}a) shows the CMIP6 ensemble mean oxygen bias in the eastern tropical Pacific OMZ relative to observations from the WOA23 dataset \citep{garcia_world_2024}.
The models underestimate oxygen concentrations in this region on average across the ensemble, with local biases exceeding \SI{0.03}{\mol\per\cubic\meter}.
The upward correction obtained from the extrapolation to a larger set of climate models (Fig.~\ref{fig:extrapolation_omz}b) reduces this bias by bringing the oxygen concentrations closer to observations, although the corrected oxygen field still has a negative bias of up to \SI{0.02}{\mol\per\cubic\meter} in some locations (Fig.~\ref{fig:extrapolation_omz}c).
Panel \ref{fig:extrapolation_omz}d) shows the effect of this correction in terms of the oxygen concentration averaged over the whole eastern tropical Pacific OMZ (defined here as the region in the eastern Pacific where the climatological observed $\mathrm{O}_2 < \SI{0.1}{\mol\per\cubic\meter}$).
This regional average view shows that the regional oxygen generated from the smaller model ensemble with biogeochemistry is very close to the directly simulated oxygen fields (orange and green curves in Fig.~\ref{fig:extrapolation_omz}d with median values of 138 and $\SI{133}{\mol\per\square\meter}$, respectively), but the regional mean oxygen distribution after applying the correction is shifted to higher values (blue curve in Fig.~\ref{fig:extrapolation_omz}d).
The resulting corrected ensemble median value of $\SI{153}{\mol\per\square\meter}$ is much closer to the true observation-based value of the Pacific OMZ oxygen concentration of $\SI{149\pm3}{\mol\per\square\meter}$ in the WOA23 dataset.
Removing the two outlier models CESM2-FV2 and E3SM-1-0 with high oxygen values (shown on Fig.~\ref{fig:extrapolation_omz}d) results in a median value of $\SI{149}{\mol\per\square\meter}$ which exactly agrees with the WOA23 observational value, but note that the observational value from the GOBAI-O2 dataset is $\SI{145}{\mol\per\square\meter}$ in this region.

For the diffusion models trained on monthly mean data and data on isopycnal layers, the multi-model mean oxygen value across the larger ensemble is closer to the observational true value in all cases except for the annual mean isopycnal setup, which we have previously seen to yield worse predictions (Fig.~\ref{appfig:extrapolation_omz}).

Given that the oxygen correction is localized and uniform in sign, we test whether the temperature and salinity predictor fields in the extra models are significantly different from those of the prior set of 14 models.
Indeed, we find that the extra models are both \SI{0.2}{\kelvin} colder and 0.05 psu fresher on average in the Pacific OMZ region (Fig.~\ref{appfig:correction_cause}), with a coherent sign at every grid cell inside the OMZ, which is consistent with higher oxygen solubility.

We have previously seen that the eastern tropical Pacific is also the region with the largest absolute extrapolation error in the leave-one-model-out setting (Fig.~\ref{fig:lomo_error}b).
However, this does not undermine the significance of the correction found in Figs.~\ref{fig:extrapolation} and \ref{fig:extrapolation_omz}: in fact, the ensemble mean signed error in the region is much smaller than the absolute error (Fig.~\ref{fig:lomo_error}a) since errors from different models do not have a coherent sign there (Fig.~\ref{fig:lomo_error}c).
The upward correction in the eastern Pacific OMZ is thus not simply an artifact of a positive LOMO error.
This is especially the case in locations with the largest upward correction: in the top 25\% of OMZ grid cells in terms of correction magnitude, the signed LOMO error only accounts for 8\% of the correction, and in the top 10\% of grid cells, the LOMO error accounts for only 3.5\% of the correction.

It is important to note that the upward correction in Figs.~\ref{fig:extrapolation}e) and \ref{fig:extrapolation_omz}b) should be interpreted as a strictly physics-conditioned correction based on the distribution $p(\mathrm{O}_2 \, | \, T, \, S)$ that the diffusion model has learned from the 14 models with biogeochemistry output.
Since this set of 14 models represents an ensemble of opportunity limited in scope by the availability of training data, it is not guaranteed that this distribution covers all possible states that the larger set of 45 models may attain.
In particular, since we have already seen that the diffusion model cannot distinguish between models with identical physics but different biogeochemistry (Fig.~\ref{fig:predictor_set}), the corrected field should not be interpreted as the oxygen field that the 45 models would exhibit if they had a biogeochemistry component.

\section{Discussion}\label{sec:discussion}

A common feature of the oxygen maps generated by the diffusion model conditioned on temperature and salinity inputs is their reduced error in the Atlantic and Southern Oceans compared to the tropical OMZ regions.
In terms of percentage error, this is partly expected because the Atlantic and Southern Oceans are well ventilated and thus have higher baseline oxygen concentrations than the poorly ventilated OMZs.
However, we showed that including DIC as a predictor improves predictive skill in the OMZs (Fig.~\ref{fig:predictor_set}).
This indicates a second reason for this common error pattern: the hydrography of the Atlantic and Southern Oceans is strongly dictated by their respective basin-scale circulation features (specifically, the Atlantic Meridional Overturning Circulation and the wind-driven \enquote{upper cell} of the Southern Ocean overturning circulation), which are well captured by large-scale fields of physical tracers.
On the other hand, the tropical OMZ oxygen inventory is influenced by a more complex interplay between circulation and biology \citep[e.g.,][]{shigemitsu_one_2017,busecke_diverging_2022,ditkovsky_unifying_2025}, which is why a biogeochemical tracer such as carbon improves reconstruction.

The difference between using purely physical ($T$, $S$) versus biogeochemical predictors (DIC) was also seen in the case of the outlier model CanESM5-CanOE: the anomalous error for this single model could only be reduced after including DIC as a conditioning field. 
The likely explanation for this behavior is that the training set contains two models with identical physical components but a different biogeochemical model (CanESM5 and CanESM5-CanOE).
Using only physical predictors, the diffusion model cannot generate skillful oxygen samples for both models, since they differ only in their biogeochemical fields; adding DIC to the predictors resolves the issue.

Concerning the training data resolution, we have tested different combinations of temporal resolution (monthly vs. annual mean) and vertical resolution (depth-integrated vs. on 10 isopycnal/density layers).
For both in-sample and out-of-sample generation tasks, we have found that using monthly data increases the diffusion model's skill compared to the annual mean data.
However, the use of data on density layers only provides enhanced skill when paired with monthly data---the annual mean data on density layers showed the largest error out of the four resolution combinations.
This is plausibly explained by the fact that the density stratification in the upper ocean contains a strong seasonal signal which can only be resolved by monthly data, while annual mean fields on fixed density layers provide a worse representation than a simple depth-integrated view of the tracer.
However, a test on historical data indicates that the annual approach accurately captures both the long-term trend and the inter-member spread in single-model, multi-member ensembles (Fig.~\ref{appfig:smile}), demonstrating its applicability to internal variability. 

Although the diffusion model has shown considerable skill in generating correct oxygen samples, including in comparison to simpler baseline models (Section \ref{sec:baselines_results}), it has several limitations.
Since the diffusion model is trained on a dataset of climate model outputs, it can not be expected to outperform the underlying climate models in terms of bias relative to observations.
However, this was not the goal of this study: instead, the diffusion model is simply an efficient method to obtain a probabilistic representation of the full multi-model ensemble's distribution, including any biases present in the training data.
In future studies, this representation could be augmented by including observational or observation-based data in the training set, thereby reducing the overall biases in the generated samples relative to the real-world ground truth.
The training dataset represents an ensemble of opportunity given the availability of model outputs on the ESGF archive (Table \ref{tab:cmip6_models}; \cite{esgf_cmip6}), but we found that the unequal number of ensemble members available for each model does not significantly affect the skill of the trained diffusion models compared to a situation with a uniform number of ensemble members per model (Fig.~\ref{appfig:composition_sensitivity}).

As diffusion models continue to gain popularity in probabilistic weather forecasting and emulation \citep[e.g.,][]{price_gencast_2024,bouabid_scorebased_2026}, future studies should further explore their use in the multi-model ensemble setting.
An obvious extension of our methodology would be to consider other biogeochemical variables, such as carbon, nutrients or the ideal age field, which are only produced by some models and can be expected to have statistical relationships to physical predictor fields.

Another natural use of learned conditional distributions such as $p(\mathrm{O}_2 \, | \, T, \, S, \, \ldots)$ is their application in emergent constraints \citep{eyring_taking_2019}, where such inter-model relationships between predictors and target are exploited using observations of the predictors.
Just as in traditional emergent constraints, such diffusion model-based constraints would need to take care in assessing potential issues related to model interdependence and spurious relationships \citep[e.g.,][]{caldwell_statistical_2014} before presenting inferred estimates.

While we have trained our diffusion models strictly on climate model outputs, it would be possible to train similar models on a mixture of climate model outputs and observational data.
This could potentially combine the accuracy of observations with the spatial covariance structure and complete coverage of model outputs, thus producing a hybrid observation-model product to complement purely data-based products such as GOBAI-O2 \cite{sharp_gobaio2_2023}.

Given the capability of diffusion models to provide probabilistic estimates including local uncertainties in tracer values, there is also the prospect of supporting observing system design by identifying locations where new observations would optimally reduce oxygen uncertainties given the existing coverage of $T$, $S$ and O$_2$ observations.

More generally, the learned conditional distribution $p(\mathrm{O}_2 \, | \, T, \, S, \, \ldots)$ could also be used to answer counterfactual questions by artificially changing the values of the input variables and interpreting the resulting changes in terms of physical and biogeochemical processes, such as the solubility-driven deoxygenation caused by an artificially imposed warming signal.

With the increasing amounts of model data being generated and archived for public use, these approaches can provide a complementary method to work with ensemble distributions in a probabilistic manner and to address questions in climate science in a new way. 

\clearpage
\acknowledgments

This research received support through Schmidt Sciences, LLC under the InMOS and M$^2$LInES projects.
This work was supported in part through the NYU IT High Performance Computing resources, services, and staff expertise.
To access and process the model outputs, this study benefited from the IPSL Data and Computing Center ESPRI, which is supported by CNRS, SU, CNES, and \'Ecole Polytechnique.
We acknowledge the World Climate Research Programme, which, through its Working Group on Coupled Modelling, coordinated and promoted CMIP6.
We thank the climate modeling groups for producing and making available their model output, the Earth System Grid Federation (ESGF) for archiving the data and providing access, and the multiple funding agencies who support CMIP6 and ESGF.
The U-Net schematic in Fig.~\ref{fig:schematic}a was created using the PlotNeuralNet library \citep{iqbal_harisiqbal88_2018}.

\datastatement

The code used for model training and to create the main figures in this manuscript is available at \url{https://doi.org/10.5281/zenodo.22963862}.
Due to the large volume of data, the full training dataset of preprocessed CMIP6 outputs can not be made available, but all CMIP6 model outputs are freely available at \url{https://esgf-node.llnl.gov/projects/cmip6/} \citep{esgf_cmip6}.

\FloatBarrier
\appendix

\begin{table*}[t]
\centering
\small
\caption{
CMIP6 model output used in this study.
Each entry gives the number of ensemble members from a particular model for which all variables in a given combination are available in the 1pctCO2 and abrupt-4xCO2 scenarios.
Note that dissolved inorganic carbon (DIC) is available for all members where oxygen is available.}
\label{tab:cmip6_models}
\begin{tabular}{lrr|@{\hspace{2em}}lrr}
\hline
 & \multicolumn{2}{c}{Ensemble members} &  & \multicolumn{2}{c}{Ensemble members} \\
\hline
Model & $T$+$S$ & $T$+$S$+O$_2$ & Model & $T$+$S$ & $T$+$S$+O$_2$ \\
\hline
ACCESS-CM2 & 1/1 & 0/0 & GFDL-CM4 & 1/1 & 1/1 \\
ACCESS-ESM1-5 & 1/2 & 1/2 & GFDL-ESM4 & 1/1 & 1/1 \\
BCC-CSM2-MR & 1/1 & 0/0 & GISS-E2-1-G & 2/4 & 0/0 \\
BCC-ESM1 & 1/1 & 0/0 & GISS-E2-1-H & 1/2 & 0/0 \\
CAMS-CSM1-0 & 2/2 & 0/0 & GISS-E2-2-G & 1/1 & 0/0 \\
CanESM5 & 6/2 & 6/2 & HadGEM3-GC31-LL & 4/1 & 0/0 \\
CanESM5-CanOE & 1/0 & 1/0 & INM-CM4-8 & 1/1 & 0/0 \\
CAS-ESM2-0 & 1/1 & 0/0 & INM-CM5-0 & 1/1 & 0/0 \\
CESM2 & 1/1 & 0/0 & IPSL-CM5A2-INCA & 1/1 & 1/1 \\
CESM2-FV2 & 1/1 & 0/0 & IPSL-CM6A-LR & 1/1 & 1/1 \\
CESM2-WACCM & 1/1 & 0/0 & IPSL-CM6A-LR-INCA & 0/1 & 0/1 \\
CESM2-WACCM-FV2 & 1/1 & 0/0 & IPSL-CM6A-MR1 & 1/1 & 0/1 \\
CIESM & 1/0 & 0/0 & MCM-UA-1-0 & 1/1 & 0/0 \\
CMCC-CM2-SR5 & 1/1 & 0/0 & MIROC-ES2L & 1/1 & 1/1 \\
CMCC-ESM2 & 1/1 & 1/0 & MIROC6 & 1/1 & 0/0 \\
CNRM-CM6-1 & 1/1 & 0/0 & MPI-ESM-1-2-HAM & 1/1 & 1/1 \\
CNRM-ESM2-1 & 10/3 & 10/3 & MPI-ESM1-2-HR & 1/1 & 1/1 \\
E3SM-1-0 & 1/0 & 0/0 & MPI-ESM1-2-LR & 1/1 & 1/1 \\
EC-Earth3 & 1/2 & 0/0 & MRI-ESM2-0 & 2/14 & 1/1 \\
EC-Earth3-AerChem & 1/1 & 0/0 & NESM3 & 1/1 & 0/0 \\
EC-Earth3-CC & 1/0 & 0/0 & NorCPM1 & 1/1 & 0/0 \\
EC-Earth3-Veg & 1/1 & 0/0 & NorESM2-LM & 1/1 & 1/1 \\
EC-Earth3-Veg-LR & 1/0 & 0/0 & NorESM2-MM & 1/0 & 0/0 \\
FGOALS-f3-L & 3/3 & 0/0 & SAM0-UNICON & 1/1 & 0/0 \\
FGOALS-g3 & 1/0 & 0/0 & UKESM1-0-LL & 4/1 & 4/1 \\
FIO-ESM-2-0 & 2/3 & 0/0 &  &  &  \\
\hline
\end{tabular}
\end{table*}

\begin{table*}[t]
\centering
\small
\caption{Layer table of the conditional ResUNet (Fig.~\ref{fig:schematic}b).
Resolutions are lat$\times$lon, and $N$ is the number of predictor channels ($N=2$ for depth-integrated $T$+$S$).
Skip connections are element-wise additions, not concatenations. The network has 2.49~M parameters.}
\label{tab:unet_layers}
\begin{tabular}{l|l|r|r}
\hline
Stage & Operation & Resolution & Channels \\
\hline
Input & concat($\bm{x}_t$, $\bm{c}$) & $180\times360$ & $1+N$ \\
Encoder 1 & $2\times$ ResBlock & $180\times360$ & 32 \\
 & AvgPool $\downarrow2$ & $90\times180$ & 32 \\
Encoder 2 & $2\times$ ResBlock & $90\times180$ & 64 \\
 & AvgPool $\downarrow2$ & $45\times90$ & 64 \\
Encoder 3 & $2\times$ ResBlock & $45\times90$ & 128 \\
 & AvgPool $\downarrow2$ & $22\times45$ & 128 \\
Bottleneck & $2\times$ ResBlock & $22\times45$ & 128 \\
Decoder 3 & bilinear $\uparrow2$ & $44\times90$ $\rightarrow$ $45\times90$ & 128 \\
 & $+$ skip (encoder 3), $2\times$ ResBlock & $45\times90$ & 128 \\
Decoder 2 & bilinear $\uparrow2$, $1\times1$ conv & $90\times180$ & 64 \\
 & $+$ skip (encoder 2), $2\times$ ResBlock & $90\times180$ & 64 \\
Decoder 1 & bilinear $\uparrow2$, $1\times1$ conv & $180\times360$ & 32 \\
 & $+$ skip (encoder 1), $2\times$ ResBlock & $180\times360$ & 32 \\
Output & $1\times1$ conv & $180\times360$ & 1 \\
\hline
\end{tabular}
\end{table*}

\begin{figure*}[t]
    \centering
    \includegraphics[width=0.8\textwidth]{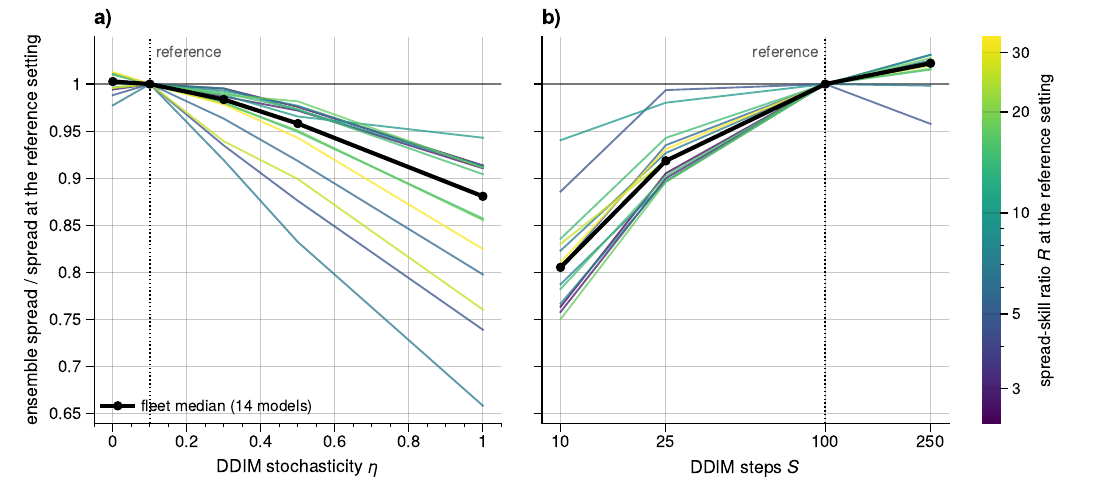}\\
    \caption{
    Sensitivity of generated ensemble spread to DDIM sampling parameters.
    \textbf{a)}, dependence of ensemble spread on the stochasticity parameter $\eta$ relative to the chosen value of $\eta=0.1$.
    \textbf{b)}, dependence of ensemble spread on the number of sampling steps $S$ relative to the chosen value of $S=100$.
    In both panels, each line represents 50 global oxygen samples drawn for one of 14 CMIP6 models, with that model being held out from training, using annual depth-integrated $T$ and $S$ predictors, evaluated over the held-out years 81--100 of the 1pctCO2 scenario.
    The lines are colored according to their spread-skill ratio $R$, with $R=1$ for perfectly calibrated spread.
    }\label{appfig:sampler_sweep}
\end{figure*}

\begin{figure*}[t]
    \centering
    \includegraphics[width=0.6\textwidth]{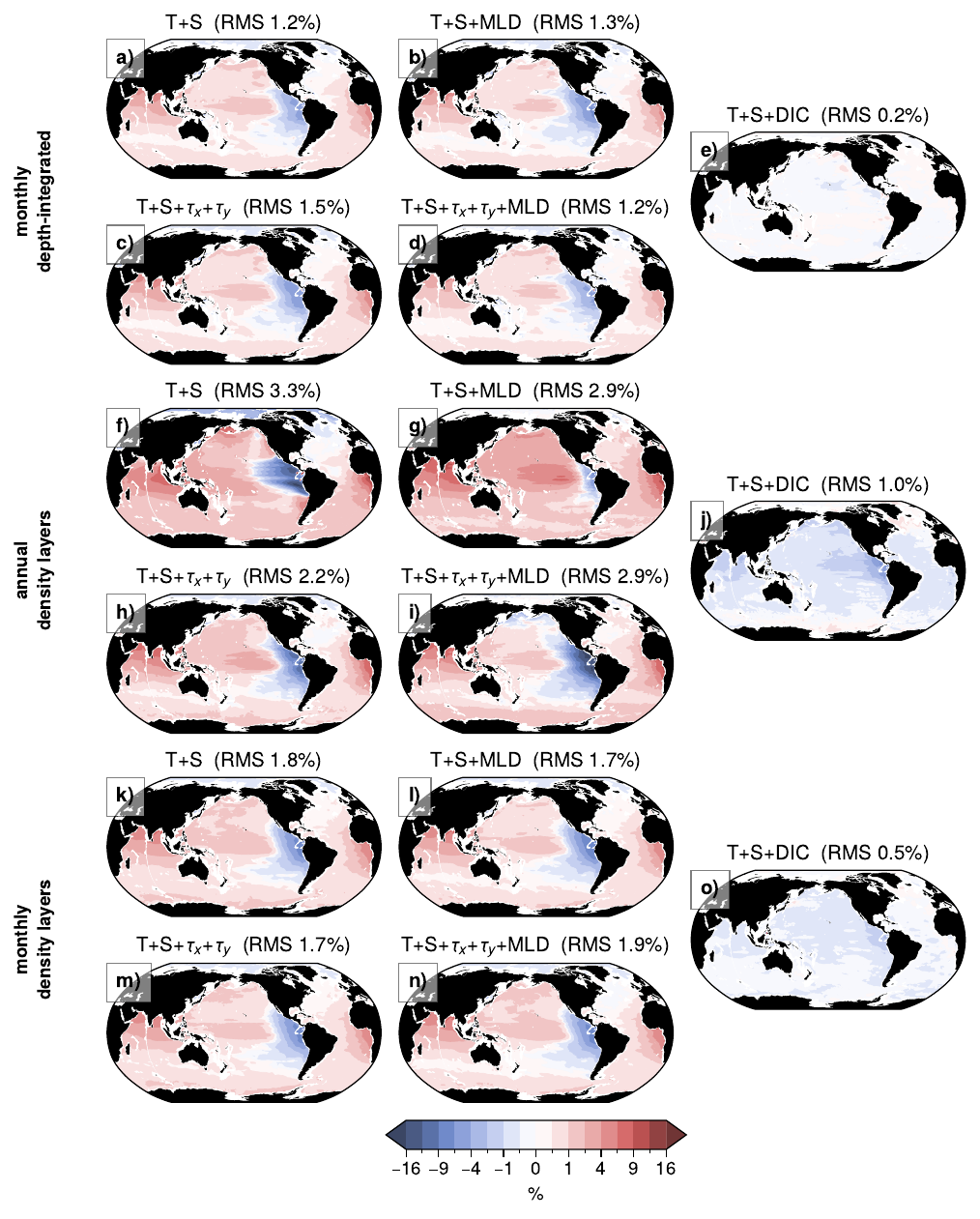}\\
    \caption{
    Sensitivity of local generative skill to the choice of predictor variables.
    Same as Fig.~\ref{fig:predictor_set} but for diffusion models trained on monthly data and/or data on isopycnal layers.
     The maps show the signed relative error across the climate model ensemble averaged over the held-out years 81-100 of the 1pctCO2 scenario.
    The numbers in parentheses state the global root mean squared (RMS) error for each panel.
    }\label{appfig:predictor_set}
\end{figure*}

\begin{figure*}[t]
    \centering
    \includegraphics[width=0.6\textwidth]{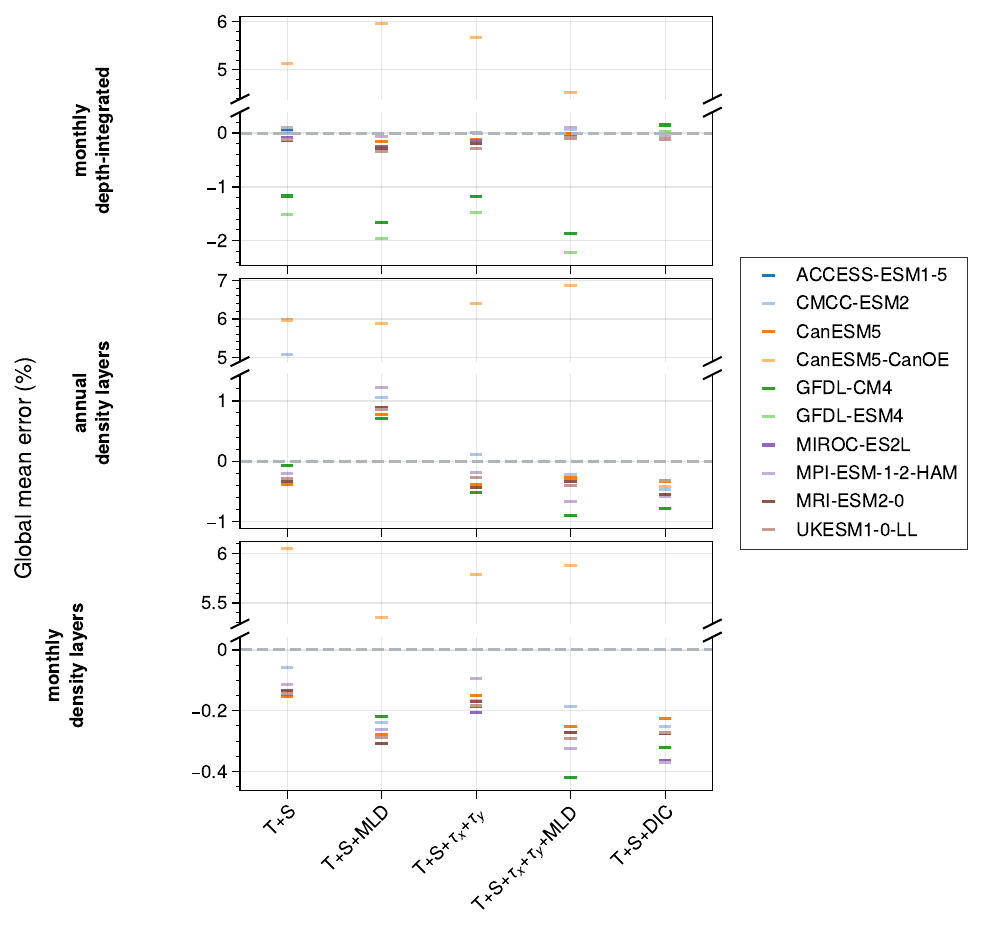}\\
    \caption{
    Sensitivity of global mean generative skill to the choice of predictor variables.
    Same as Fig.~\ref{fig:predictor_boxplot} but for diffusion models trained on monthly data and/or data on isopycnal layers.
    The data and variable combinations are as in Fig.~\ref{fig:predictor_set}, but the colored lines show the area-weighted global mean relative oxygen prediction error for each climate model separately.
    The y-axes are cut to enhance the detail for the groups of models with errors close to 0\%.
    }\label{appfig:predictor_boxplot}
\end{figure*}

\begin{figure*}[t]
    \centering
    \includegraphics[width=\textwidth]{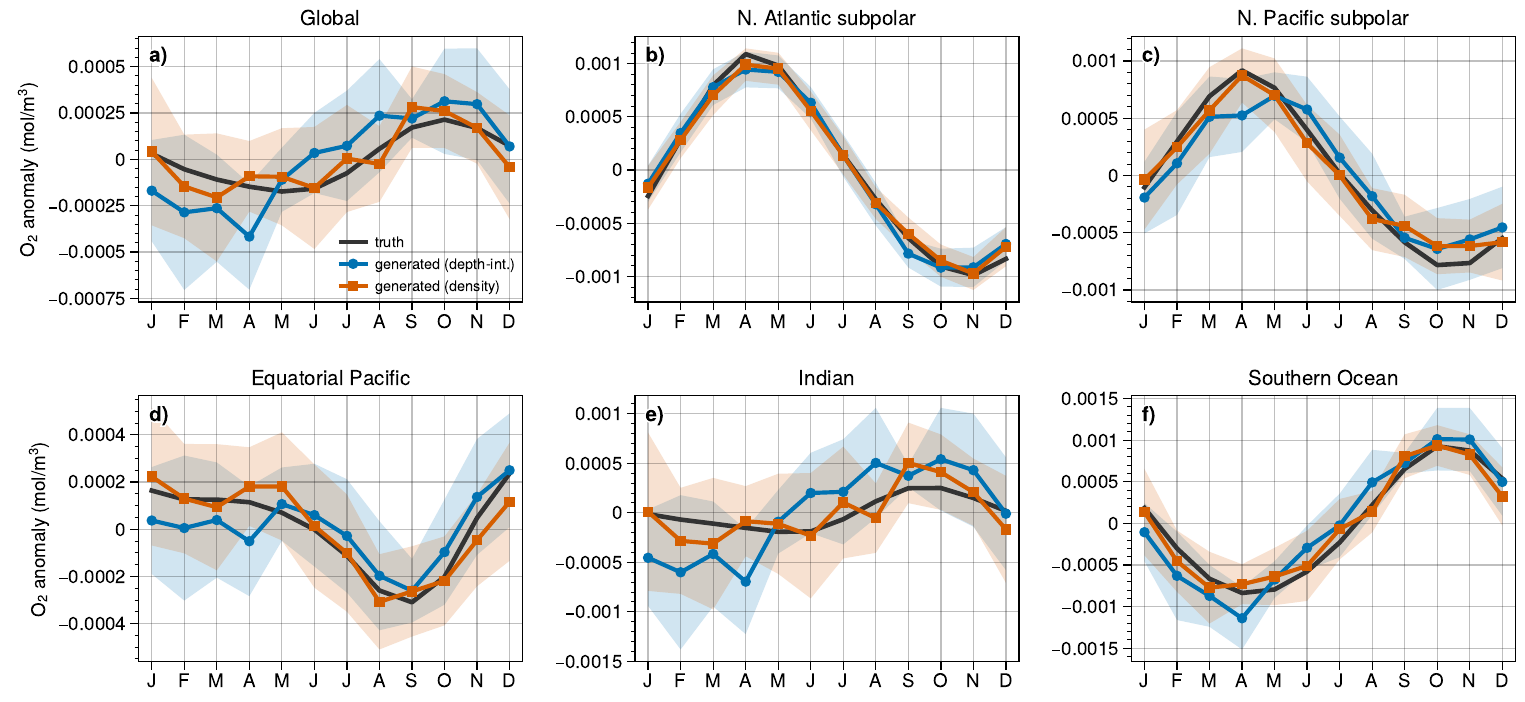}\\
    \caption{
    Seasonal cycle of true and generated regional oxygen fields using monthly training data.
    Each panel includes the multi-model mean ground truth seasonal cycle (black), and the multi-model mean generated seasonal cycle using diffusion models trained on depth-integrated monthly data (blue) and monthly data on isopycnal layers (orange).
    The shading shows one standard deviation across the ensemble.
    The annual mean bias in each panel is subtracted to emphasize amplitude and phase of the true and generated seasonal cycles.
    }\label{appfig:seasonal_cycle_anomaly}
\end{figure*}

\begin{figure*}[t]
    \centering
    \includegraphics[width=0.55\textwidth]{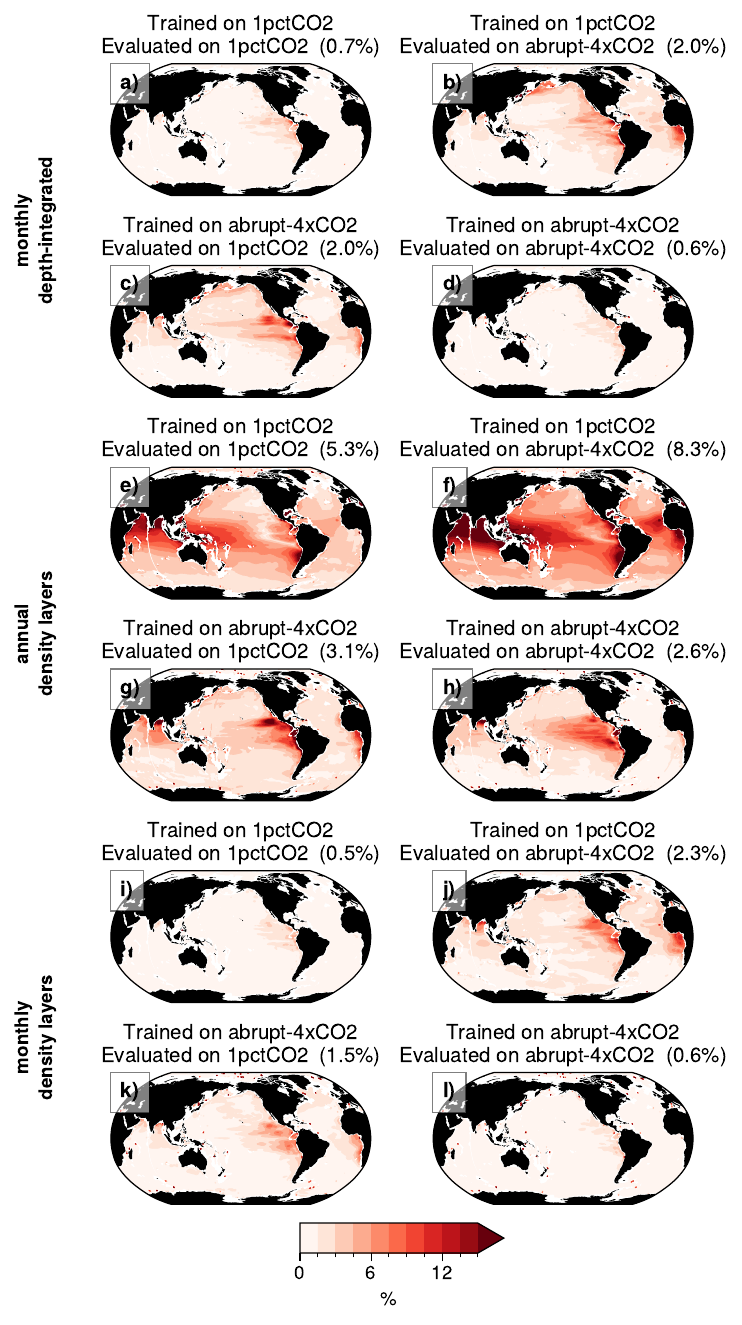}\\
    \caption{
    Generative skill in applications to out-of-sample forcing scenarios.
    Same as Fig.~\ref{fig:scenario_transfer} but for diffusion models trained on monthly data and/or data on isopycnal layers.
    All maps show the mean absolute percentage error (MAPE) across the climate model ensemble in years 81--100, for a diffusion model trained and evaluated on outputs for specific scenarios (1pctCO2 and abrupt-4xCO2).
    \textbf{a, e, i)}, trained and evaluated on 1pctCO2 (in-sample).
    \textbf{b, f, j)}, trained on 1pctCO2 and evaluated on abrupt-4xCO2 (out-of-sample).
    \textbf{c, g, k)}, trained on abrupt-4xCO2 and evaluated on 1pctCO2 (out-of-sample).
    \textbf{d, h, l)}, trained and evaluated on abrupt-4xCO2 (in-sample).
    The numbers in parentheses state the global mean MAPE.
    }\label{appfig:scenario_transfer}
\end{figure*}

\begin{figure*}[t]
    \centering
    \includegraphics[width=\textwidth]{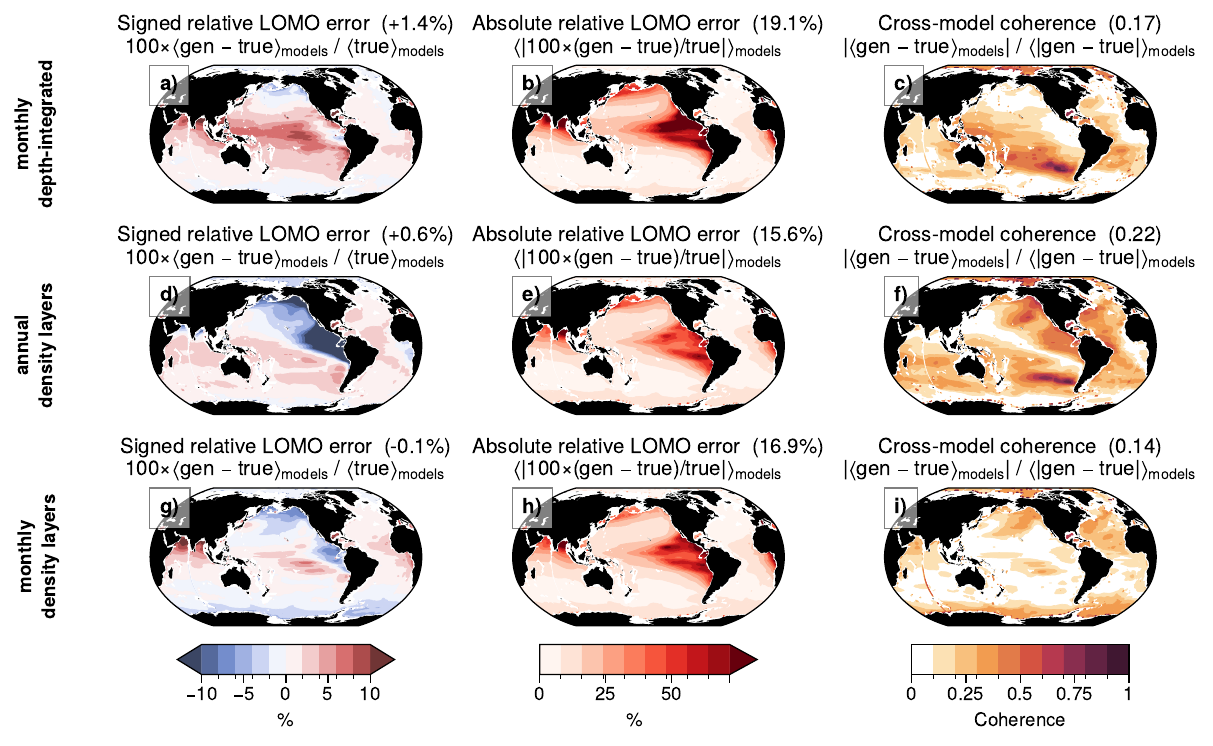}\\
    \caption{
    Generative skill in applications to out-of-sample climate models (for annual mean depth-integrated fields).
    Same as Fig.~\ref{fig:lomo_error} but for diffusion models trained on monthly data and/or data on isopycnal layers.
    Left column, mean absolute value across the climate model ensemble of the oxygen prediction error for each climate model in years 81--100 of the 1pctCO2 scenario, for a diffusion model trained on outputs from all other climate models.
    Right column, same as left column but relative to the ground truth oxygen field.
    }\label{appfig:lomo_error}
\end{figure*}

\begin{figure*}[t]
    \centering
    \includegraphics[width=0.7\textwidth]{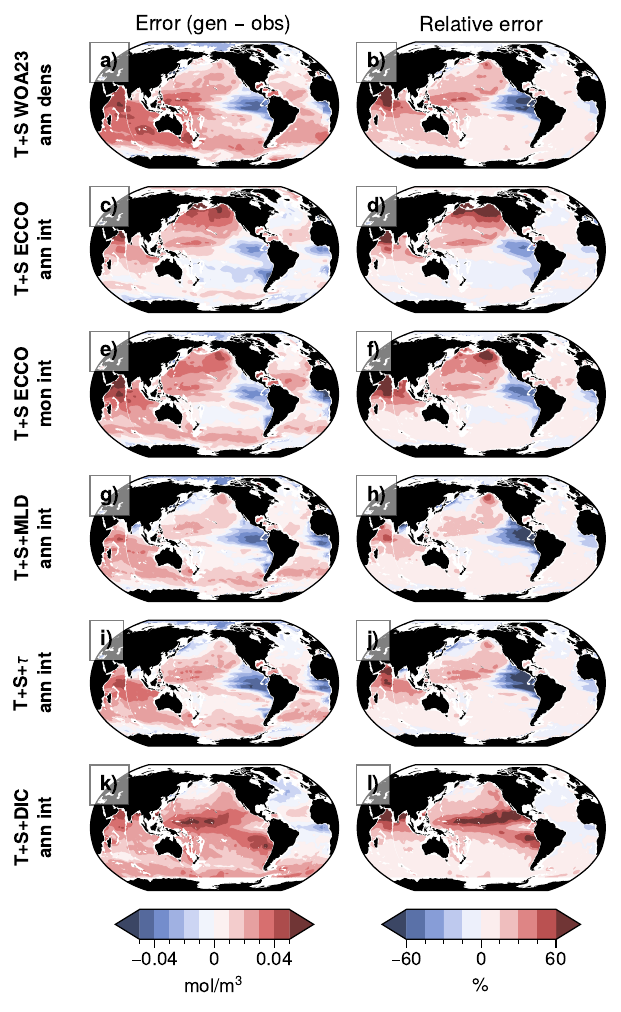}\\
    \caption{
    Observational validation of learned generative conditional distribution.
    Same as Fig.~\ref{fig:obs_validation}d) but for diffusion models trained on monthly data and/or data on isopycnal layers.
    The maps show the relative oxygen prediction error in percent.
    }\label{appfig:obs_validation}
\end{figure*}

\begin{figure*}[t]
    \centering
    \includegraphics[width=0.9\textwidth]{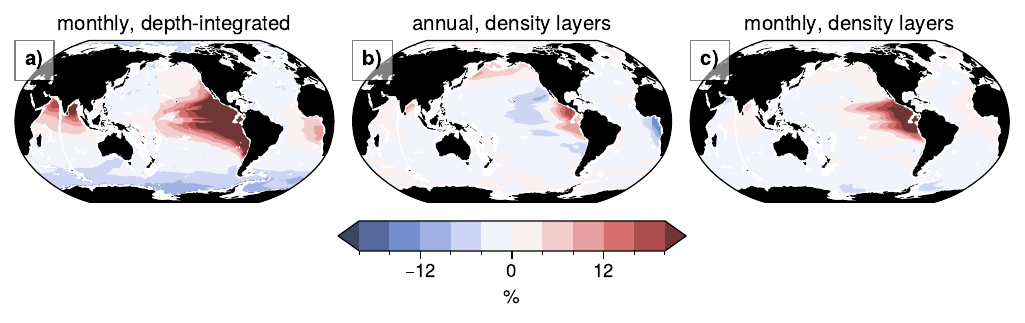}\\
    \caption{
    Oxygen generation based on physical predictors for a larger multi-model ensemble.
    Same as Fig.~\ref{fig:extrapolation}e) but for diffusion models trained on monthly data and/or data on isopycnal layers.
    The maps show relative difference in ensemble mean oxygen field predicted for the larger set of models vs. the field predicted for the set of models with ground truth oxygen fields available.
    }\label{appfig:extrapolation}
\end{figure*}

\begin{figure*}[t]
    \centering
    \includegraphics[width=0.95\textwidth]{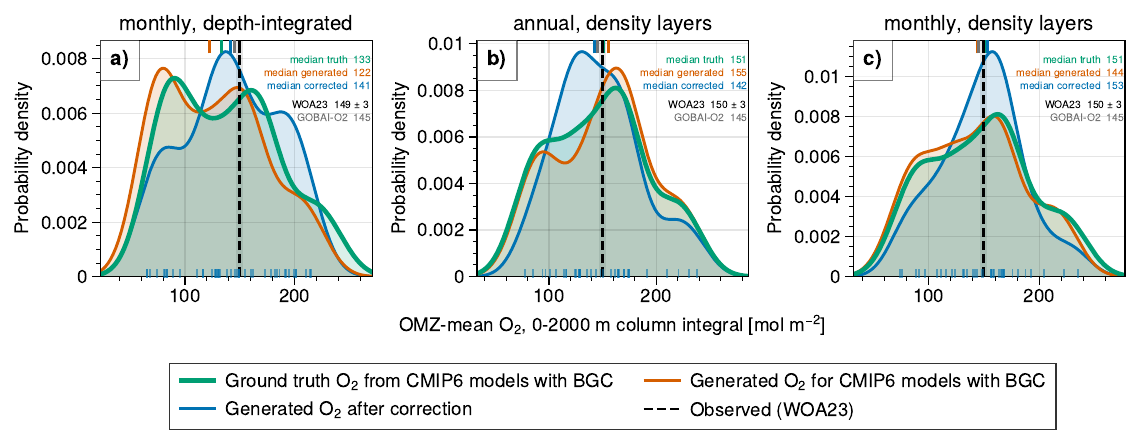}\\
    \caption{
    Implications of oxygen extrapolation to larger ensemble size on the eastern tropical Pacific Oxygen Minimum Zone.
    Same as Fig.~\ref{fig:extrapolation_omz}d) but for diffusion models trained on monthly data and/or data on isopycnal layers.
    The panels show the inter-model distribution of eastern tropical Pacific oxygen concentration before and after applying the correction.
    The curves are estimated from histograms of individual model values via Gaussian kernel density estimation, with a fixed kernel bandwidth of 0.4 times each curve's own standard deviation.
    The OMZ is defined as the area within 30° of the equator in the Pacific whose observed 0–2000 m column-integrated oxygen is below \SI{200}{\mol\per\square\meter}.
    The dashed vertical line indicates the observational value from the World Ocean Atlas 2023 \citep{garcia_world_2024}, and the WOA23 uncertainty stated in the legend is its standard error averaged over the region assuming fully correlated errors.
    The short blue lines at the bottom of the panel indicates individual model values including two highlighted outlier models (CESM2-FV2 and E3SM-1-0).
    The short colored lines at the top of the panel indicates the median of each distribution.
    }\label{appfig:extrapolation_omz}
\end{figure*}

\begin{figure*}[t]
    \centering
    \includegraphics[width=\textwidth]{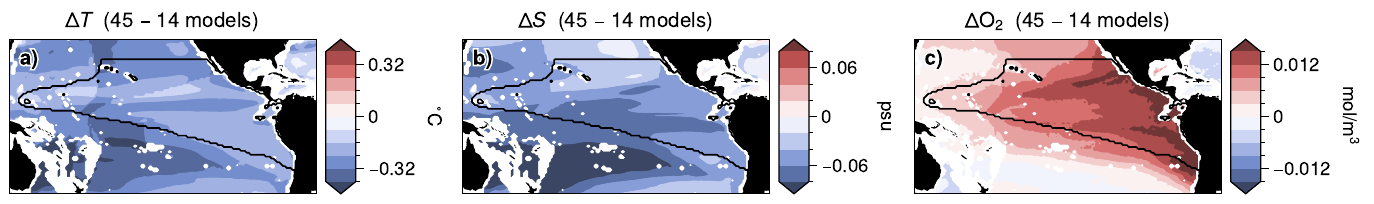}\\
    \caption{
    Composite difference in depth-integrated $T$, $S$, and O$_2$ between the set of 45 models used for extrapolation in Section \ref{sec:extrapolation} and the set of 14 models which carry an oxygen variable.
    The $T$ and $S$ fields are taken directly from model output, while the O$_2$ field in panel c) is predicted by the diffusion model, as described in Section \ref{fig:extrapolation}.
    }\label{appfig:correction_cause}
\end{figure*}

\begin{figure*}[t]
    \centering
    \includegraphics[width=0.95\textwidth]{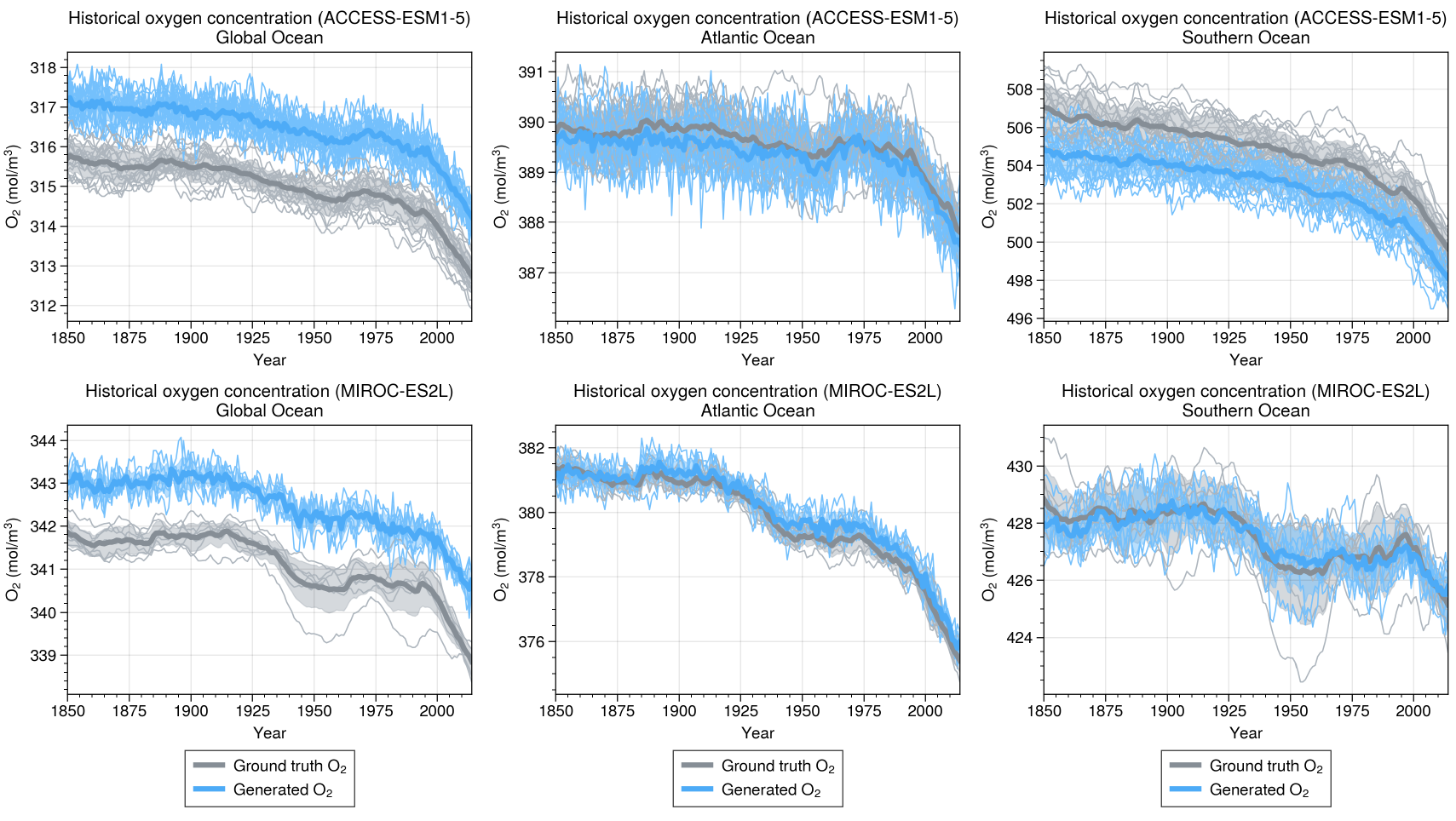}\\
    \caption{
    Historical time series of basin-mean oxygen concentration in two single-model multi-member ensembles.
    First row \textbf{a)}--\textbf{c)}: ground truth and generated oxygen for 19 members of the ACCESS-ESM1-5 model, in the global (a), Atlantic (b), and Southern Ocean (c).
    Second row \textbf{d)}--\textbf{f)}: ground truth and generated oxygen for 10 members of the MIROC-ES2L model in the same regions.
    }\label{appfig:smile}
\end{figure*}

\begin{figure*}[t]
    \centering
    \includegraphics[width=0.7\textwidth]{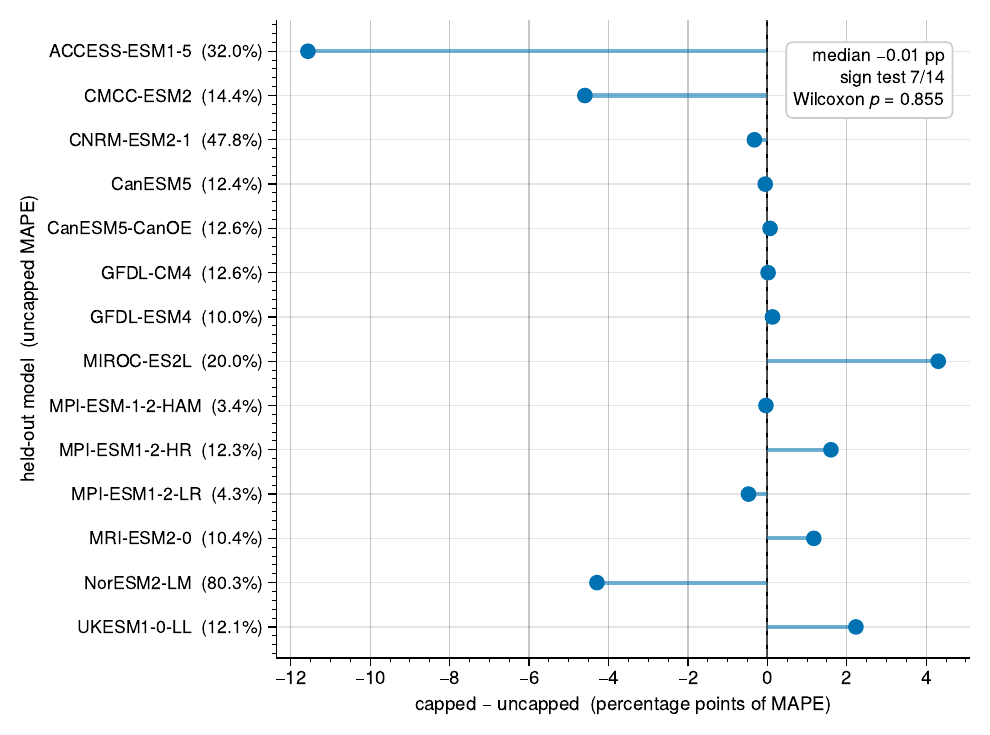}\\
    \caption{
    Sensitivity of diffusion model skill to capping the number of ensemble members per climate model.
    The points show the change in MAPE when using only one ensemble member for each model (\enquote{capped}) versus the default case of using all available ensemble members for each model, even when those numbers are unequal across different models (\enquote{uncapped}).
    The ensemble median difference is 0.01 percentage points, with $p=0.855$ under a Wilcoxon signed-rank test.
    }\label{appfig:composition_sensitivity}
\end{figure*}

\clearpage
\bibliographystyle{ametsocV6}
\bibliography{diffusion_oxygen}

\begin{thebibliography}{67}
\providecommand{\natexlab}[1]{#1}
\providecommand{\url}[1]{\texttt{#1}}
\renewcommand{\UrlFont}{\rmfamily}
\providecommand{\urlprefix}{URL }
\expandafter\ifx\csname urlstyle\endcsname\relax
  \providecommand{\doi}[1]{https://doi.org/\discretionary{}{}{}#1}\else
  \providecommand{\doi}{https://doi.org/\discretionary{}{}{}\begingroup
  \urlstyle{rm}\Url}\fi
\providecommand{\eprint}[2][]{\url{#2}}

\bibitem[{Aich et~al.(2026)Aich, Hess, Pan, Bathiany, Huang,, and
  Boers}]{aich_conditional_2026}
Aich, M., P.~Hess, B.~Pan, S.~Bathiany, Y.~Huang, and N.~Boers, 2026:
  arXiv:2404.14416. Conditional diffusion models for downscaling and bias
  correction of {{Earth}} system model precipitation. arXiv,
  \doi{10.48550/arXiv.2404.14416}, \eprint{2404.14416}.

\bibitem[{Andrews et~al.(2017)Andrews, Buitenhuis, Le~Qu{\'e}r{\'e},, and
  Suntharalingam}]{andrews_biogeochemical_2017}
Andrews, O., E.~Buitenhuis, C.~Le~Qu{\'e}r{\'e}, and P.~Suntharalingam, 2017:
  Biogeochemical modelling of dissolved oxygen in a changing ocean.
  \textit{Philosophical Transactions of the Royal Society A: Mathematical,
  Physical and Engineering Sciences}, \textbf{375~(2102)}, 20160\,328,
  \doi{10.1098/rsta.2016.0328}.

\bibitem[{Bassetti et~al.(2024)Bassetti, Hutchinson, Tebaldi,, and
  Kravitz}]{bassetti_diffesm_2024}
Bassetti, S., B.~Hutchinson, C.~Tebaldi, and B.~Kravitz, 2024: {{DiffESM}}:
  {{Conditional Emulation}} of {{Temperature}} and {{Precipitation}} in {{Earth
  System Models With 3D Diffusion Models}}. \textit{Journal of Advances in
  Modeling Earth Systems}, \textbf{16~(10)}, e2023MS004\,194,
  \doi{10.1029/2023MS004194}.

\bibitem[{Beucler et~al.(2024)}]{beucler_climateinvariant_2024}
Beucler, T., and Coauthors, 2024: Climate-invariant machine learning.
  \textit{Science Advances}, \textbf{10~(6)}, eadj7250,
  \doi{10.1126/sciadv.adj7250}.

\bibitem[{Bopp et~al.(2013)}]{bopp_multiple_2013}
Bopp, L., and Coauthors, 2013: Multiple stressors of ocean ecosystems in the
  21st century: Projections with {{CMIP5}} models. \textit{Biogeosciences},
  \textbf{10~(10)}, 6225--6245, \doi{10.5194/bg-10-6225-2013}.

\bibitem[{Bouabid et~al.(2026)Bouabid, Souza,, and
  Ferrari}]{bouabid_scorebased_2026}
Bouabid, S., A.~N. Souza, and R.~Ferrari, 2026: Score-{{Based Generative
  Emulation}} of {{Impact-Relevant Earth System Model Outputs}}.
  \textit{Journal of Advances in Modeling Earth Systems}, \textbf{18~(3)},
  e2025MS005\,558, \doi{10.1029/2025MS005558}.

\bibitem[{Boyd et~al.(2015)Boyd, Lennartz, Glover,, and
  Doney}]{boyd_biological_2015}
Boyd, P.~W., S.~T. Lennartz, D.~M. Glover, and S.~C. Doney, 2015: Biological
  ramifications of climate-change-mediated oceanic multi-stressors.
  \textit{Nature Climate Change}, \textbf{5~(1)}, 71--79,
  \doi{10.1038/nclimate2441}.

\bibitem[{Breiman(2001)}]{breiman_random_2001}
Breiman, L., 2001: Random {{Forests}}. \textit{Machine Learning},
  \textbf{45~(1)}, 5--32, \doi{10.1023/A:1010933404324}.

\bibitem[{Breitburg et~al.(2018)}]{breitburg_declining_2018}
Breitburg, D., and Coauthors, 2018: Declining oxygen in the global ocean and
  coastal waters. \textit{Science}, \textbf{359~(6371)}, eaam7240,
  \doi{10.1126/science.aam7240}.

\bibitem[{Buchanan and Tagliabue(2021)Buchanan, and
  Tagliabue}]{buchanan_regional_2021}
Buchanan, P.~J., and A.~Tagliabue, 2021: The {{Regional Importance}} of
  {{Oxygen Demand}} and {{Supply}} for {{Historical Ocean Oxygen Trends}}.
  \textit{Geophysical Research Letters}, \textbf{48~(20)}, e2021GL094\,797,
  \doi{10.1029/2021GL094797}.

\bibitem[{Busecke et~al.(2022)Busecke, Resplandy, Ditkovsky,, and
  John}]{busecke_diverging_2022}
Busecke, J. J.~M., L.~Resplandy, S.~J. Ditkovsky, and J.~G. John, 2022:
  Diverging {{Fates}} of the {{Pacific Ocean Oxygen Minimum Zone}} and {{Its
  Core}} in a {{Warming World}}. \textit{AGU Advances}, \textbf{3~(6)},
  e2021AV000\,470, \doi{10.1029/2021AV000470}.

\bibitem[{Caldwell et~al.(2014)Caldwell, Bretherton, Zelinka, Klein, Santer,,
  and Sanderson}]{caldwell_statistical_2014}
Caldwell, P.~M., C.~S. Bretherton, M.~D. Zelinka, S.~A. Klein, B.~D. Santer,
  and B.~M. Sanderson, 2014: Statistical significance of climate sensitivity
  predictors obtained by data mining. \textit{Geophysical Research Letters},
  \textbf{41~(5)}, 1803--1808, \doi{10.1002/2014GL059205}.

\bibitem[{{de Boyer Mont{\'e}gut} et~al.(2004){de Boyer Mont{\'e}gut}, Madec,
  Fischer, Lazar,, and Iudicone}]{deboyermontegut_mixed_2004}
{de Boyer Mont{\'e}gut}, C., G.~Madec, A.~S. Fischer, A.~Lazar, and
  D.~Iudicone, 2004: Mixed layer depth over the global ocean: {{An}}
  examination of profile data and a profile-based climatology. \textit{Journal
  of Geophysical Research: Oceans}, \textbf{109~(C12)},
  \doi{10.1029/2004JC002378}.

\bibitem[{Delteil et~al.(2026)Delteil, L{\'e}vy,, and
  Bopp}]{delteil_emerging_2026}
Delteil, M., M.~L{\'e}vy, and L.~Bopp, 2026: Emerging {{Climate Signals}} in
  {{Tropical Oxygen Minimum Zones}}. \textit{Biogeosciences}, \textbf{23~(7)},
  2205--2233, \doi{10.5194/bg-23-2205-2026}.

\bibitem[{Deutsch et~al.(2015)Deutsch, Ferrel, Seibel, P{\"o}rtner,, and
  Huey}]{deutsch_climate_2015}
Deutsch, C., A.~Ferrel, B.~Seibel, H.-O. P{\"o}rtner, and R.~B. Huey, 2015:
  Climate change tightens a metabolic constraint on marine habitats.
  \textit{Science}, \textbf{348~(6239)}, 1132--1135,
  \doi{10.1126/science.aaa1605}.

\bibitem[{Deutsch et~al.(2024)Deutsch, Penn,, and Lucey}]{deutsch_climate_2024}
Deutsch, C., J.~L. Penn, and N.~Lucey, 2024: Climate, {{Oxygen}}, and the
  {{Future}} of {{Marine Biodiversity}}. \textit{Annual Review of Marine
  Science}, \textbf{16~(Volume 16, 2024)}, 217--245,
  \doi{10.1146/annurev-marine-040323-095231}.

\bibitem[{Ditkovsky and Resplandy(2025)Ditkovsky, and
  Resplandy}]{ditkovsky_unifying_2025}
Ditkovsky, S., and L.~Resplandy, 2025: Unifying {{Future Ocean Oxygen
  Projections Using}} an {{Oxygen Water Mass Framework}}. \textit{Journal of
  Geophysical Research: Oceans}, \textbf{130~(5)}, e2025JC022\,333,
  \doi{10.1029/2025JC022333}.

\bibitem[{{Earth System Grid Federation}(2026)}]{esgf_cmip6}
{Earth System Grid Federation}, 2026: {{ESGF MetaGrid}}.
  https://esgf-node.llnl.gov/projects/cmip6/.

\bibitem[{{ECCO Consortium} et~al.(2024){ECCO Consortium}, Fukumori, Wang,
  Fenty, Forget, Heimbach,, and Ponte}]{eccoconsortium_ecco_2024}
{ECCO Consortium}, I.~Fukumori, O.~Wang, I.~Fenty, G.~Forget, P.~Heimbach, and
  R.~M. Ponte, 2024: {{ECCO Central Estimate}}.

\bibitem[{Eyring et~al.(2019)}]{eyring_taking_2019}
Eyring, V., and Coauthors, 2019: Taking climate model evaluation to the next
  level. \textit{Nature Climate Change}, \textbf{9~(2)}, 102--110,
  \doi{10.1038/s41558-018-0355-y}.

\bibitem[{Forget et~al.(2015)Forget, Campin, Heimbach, Hill, Ponte,, and
  Wunsch}]{forget_ecco_2015}
Forget, G., J.-M. Campin, P.~Heimbach, C.~N. Hill, R.~M. Ponte, and C.~Wunsch,
  2015: {{ECCO}} version 4: An integrated framework for non-linear inverse
  modeling and global ocean state estimation. \textit{Geoscientific Model
  Development}, \textbf{8~(10)}, 3071--3104, \doi{10.5194/gmd-8-3071-2015}.

\bibitem[{Fr{\"o}licher et~al.(2009)Fr{\"o}licher, Joos, Plattner, Steinacher,,
  and Doney}]{frolicher_natural_2009}
Fr{\"o}licher, T.~L., F.~Joos, G.-K. Plattner, M.~Steinacher, and S.~C. Doney,
  2009: Natural variability and anthropogenic trends in oceanic oxygen in a
  coupled carbon cycle--climate model ensemble. \textit{Global Biogeochemical
  Cycles}, \textbf{23~(1)}, \doi{10.1029/2008GB003316}.

\bibitem[{Fr{\"o}licher et~al.(2016)Fr{\"o}licher, Rodgers, Stock,, and
  Cheung}]{frolicher_sources_2016}
Fr{\"o}licher, T.~L., K.~B. Rodgers, C.~A. Stock, and W.~W.~L. Cheung, 2016:
  Sources of uncertainties in 21st century projections of potential ocean
  ecosystem stressors. \textit{Global Biogeochemical Cycles}, \textbf{30~(8)},
  1224--1243, \doi{10.1002/2015GB005338}.

\bibitem[{Garcia et~al.(2024)}]{garcia_world_2024}
Garcia, H.~E., and Coauthors, 2024: World {{Ocean Atlas}} 2023, {{Volume}} 3:
  {{Dissolved Oxygen}}, {{Apparent Oxygen Utilization}}, and {{Dissolved Oxygen
  Saturation}}. \textit{NOAA Atlas NESDIS ; 91}, \doi{10.25923/RB67-NS53}.

\bibitem[{Gneiting and Raftery(2007)Gneiting, and
  Raftery}]{gneiting_strictly_2007}
Gneiting, T., and A.~E. Raftery, 2007: Strictly {{Proper Scoring Rules}},
  {{Prediction}}, and {{Estimation}}. \textit{Journal of the American
  Statistical Association}, \textbf{102~(477)}, 359--378,
  \doi{10.1198/016214506000001437}.

\bibitem[{Gruber et~al.(2021)Gruber, Boyd, Fr{\"o}licher,, and
  Vogt}]{gruber_biogeochemical_2021}
Gruber, N., P.~W. Boyd, T.~L. Fr{\"o}licher, and M.~Vogt, 2021: Biogeochemical
  extremes and compound events in the ocean. \textit{Nature},
  \textbf{600~(7889)}, 395--407, \doi{10.1038/s41586-021-03981-7}.

\bibitem[{Hersbach et~al.(2020)}]{hersbach_era5_2020}
Hersbach, H., and Coauthors, 2020: The {{ERA5}} global reanalysis.
  \textit{Quarterly Journal of the Royal Meteorological Society},
  \textbf{146~(730)}, 1999--2049, \doi{10.1002/qj.3803}.

\bibitem[{Ho et~al.(2020)Ho, Jain,, and Abbeel}]{ho_denoising_2020}
Ho, J., A.~Jain, and P.~Abbeel, 2020: arXiv:2006.11239. Denoising {{Diffusion
  Probabilistic Models}}. arXiv, \doi{10.48550/arXiv.2006.11239},
  \eprint{2006.11239}.

\bibitem[{Iqbal(2018)}]{iqbal_harisiqbal88_2018}
Iqbal, H., 2018: {{HarisIqbal88}}/{{PlotNeuralNet}} v1.0.0. Zenodo,
  \doi{10.5281/zenodo.2526396}.

\bibitem[{Ito et~al.(2024)Ito, Cervania, Cross, Ainchwar,, and
  Delawalla}]{ito_mapping_2024}
Ito, T., A.~Cervania, K.~Cross, S.~Ainchwar, and S.~Delawalla, 2024: Mapping
  {{Dissolved Oxygen Concentrations}} by {{Combining Shipboard}} and {{Argo
  Observations Using Machine Learning Algorithms}}. \textit{Journal of
  Geophysical Research: Machine Learning and Computation}, \textbf{1~(3)},
  e2024JH000\,272, \doi{10.1029/2024JH000272}.

\bibitem[{Ito et~al.(2026)}]{ito_are_2026}
Ito, T., and Coauthors, 2026: Are {{Simulated Ocean Deoxygenation Rates
  Consistent}} with the {{Observational Reconstructions}}? \textit{Annual
  Review of Earth and Planetary Sciences},
  \doi{10.1146/annurev-earth-032524-123111}.

\bibitem[{Keeling et~al.(2010)Keeling, K{\"o}rtzinger,, and
  Gruber}]{keeling_ocean_2010}
Keeling, R.~F., A.~K{\"o}rtzinger, and N.~Gruber, 2010: Ocean {{Deoxygenation}}
  in a {{Warming World}}. \textit{Annual Review of Marine Science},
  \textbf{2~(Volume 2, 2010)}, 199--229,
  \doi{10.1146/annurev.marine.010908.163855}.

\bibitem[{Keppler et~al.(2023)Keppler, Landsch{\"u}tzer, Lauvset,, and
  Gruber}]{keppler_recent_2023}
Keppler, L., P.~Landsch{\"u}tzer, S.~K. Lauvset, and N.~Gruber, 2023: Recent
  {{Trends}} and {{Variability}} in the {{Oceanic Storage}} of {{Dissolved
  Inorganic Carbon}}. \textit{Global Biogeochemical Cycles}, \textbf{37~(5)},
  e2022GB007\,677, \doi{10.1029/2022GB007677}.

\bibitem[{Kwiatkowski et~al.(2020)}]{kwiatkowski_twentyfirst_2020}
Kwiatkowski, L., and Coauthors, 2020: Twenty-first century ocean warming,
  acidification, deoxygenation, and upper-ocean nutrient and primary production
  decline from {{CMIP6}} model projections. \textit{Biogeosciences},
  \textbf{17~(13)}, 3439--3470, \doi{10.5194/bg-17-3439-2020}.

\bibitem[{Le~Grix et~al.(2025)Le~Grix, Burger,, and
  Fr{\"o}licher}]{legrix_surface_2025}
Le~Grix, N., F.~A. Burger, and T.~L. Fr{\"o}licher, 2025: Surface and
  {{Subsurface Compound Marine Heatwave}} and {{Biogeochemical Extremes Under
  Climate Change}}. \textit{Global Biogeochemical Cycles}, \textbf{39~(5)},
  e2025GB008\,514, \doi{10.1029/2025GB008514}.

\bibitem[{Li et~al.(2020)Li, Huang, Ding, Liu, Yu,, and
  Huang}]{li_increasing_2020}
Li, C., J.~Huang, L.~Ding, X.~Liu, H.~Yu, and J.~Huang, 2020: Increasing
  {{Escape}} of {{Oxygen From Oceans Under Climate Change}}.
  \textit{Geophysical Research Letters}, \textbf{47~(11)}, e2019GL086\,345,
  \doi{10.1029/2019GL086345}.

\bibitem[{Li et~al.(2024)Li, Carver, {Lopez-Gomez}, Sha,, and
  Anderson}]{li_generative_2024}
Li, L., R.~Carver, I.~{Lopez-Gomez}, F.~Sha, and J.~Anderson, 2024: Generative
  emulation of weather forecast ensembles with diffusion models.
  \textit{Science Advances}, \textbf{10~(13)}, eadk4489,
  \doi{10.1126/sciadv.adk4489}.

\bibitem[{Limburg et~al.(2020)Limburg, Breitburg, Swaney,, and
  Jacinto}]{limburg_ocean_2020}
Limburg, K.~E., D.~Breitburg, D.~P. Swaney, and G.~Jacinto, 2020: Ocean
  {{Deoxygenation}}: {{A Primer}}. \textit{One Earth}, \textbf{2~(1)}, 24--29,
  \doi{10.1016/j.oneear.2020.01.001}.

\bibitem[{Ling et~al.(2024)}]{ling_diffusion_2024}
Ling, F., and Coauthors, 2024: Diffusion model-based probabilistic downscaling
  for 180-year {{East Asian}} climate reconstruction. \textit{npj Climate and
  Atmospheric Science}, \textbf{7~(1)}, 131, \doi{10.1038/s41612-024-00679-1}.

\bibitem[{Long et~al.(2019)Long, Ito,, and Deutsch}]{long_oxygen_2019}
Long, M.~C., T.~Ito, and C.~Deutsch, 2019: Oxygen projections for the future.
  \textit{Ocean deoxygenation: Everyone's problem}, 171--212.

\bibitem[{{Lopez-Gomez} et~al.(2025){Lopez-Gomez}, Wan, {Zepeda-N{\'u}{\~n}ez},
  Schneider, Anderson,, and Sha}]{lopez-gomez_dynamicalgenerative_2025}
{Lopez-Gomez}, I., Z.~Y. Wan, L.~{Zepeda-N{\'u}{\~n}ez}, T.~Schneider,
  J.~Anderson, and F.~Sha, 2025: Dynamical-generative downscaling of climate
  model ensembles. \textit{Proceedings of the National Academy of Sciences},
  \textbf{122~(17)}, e2420288\,122, \doi{10.1073/pnas.2420288122}.

\bibitem[{Lu et~al.(2024)}]{lu_oxygenerator_2024}
Lu, B., and Coauthors, 2024: arXiv:2405.07233. {{OXYGENERATOR}}:
  {{Reconstructing Global Ocean Deoxygenation Over}} a {{Century}} with {{Deep
  Learning}}. arXiv, \doi{10.48550/arXiv.2405.07233}, \eprint{2405.07233}.

\bibitem[{Mardani et~al.(2025)}]{mardani_residual_2025}
Mardani, M., and Coauthors, 2025: Residual corrective diffusion modeling for
  km-scale atmospheric downscaling. \textit{Communications Earth \&
  Environment}, \textbf{6~(1)}, 124, \doi{10.1038/s43247-025-02042-5}.

\bibitem[{McDougall and Barker(2011)McDougall, and
  Barker}]{mcdougallGettingStartedTEOS102011}
McDougall, T.~J., and P.~M. Barker, 2011: \textit{Getting Started with
  {{TEOS-10}} and the {{Gibbs Seawater}} ({{GSW}}) {{Oceanographic Toolbox}}}.
  SCOR/IAPSO WG127.

\bibitem[{Meinshausen(2006)}]{meinshausen_quantile_2006}
Meinshausen, N., 2006: Quantile {{Regression Forests}}. \textit{Journal of
  Machine Learning Research}, \textbf{7~(35)}, 983--999.

\bibitem[{Meunier et~al.(2025)Meunier, Kamm, Gachon, Lguensat,, and
  Deshayes}]{meunier_learning_2025}
Meunier, E., D.~Kamm, G.~Gachon, R.~Lguensat, and J.~Deshayes, 2025:
  arXiv:2502.02499. Learning to generate physical ocean states: {{Towards}}
  hybrid climate modeling. arXiv, \doi{10.48550/arXiv.2502.02499},
  \eprint{2502.02499}.

\bibitem[{Nichol and Dhariwal(2021)Nichol, and Dhariwal}]{nichol_improved_2021}
Nichol, A., and P.~Dhariwal, 2021: arXiv:2102.09672. Improved {{Denoising
  Diffusion Probabilistic Models}}. arXiv, \doi{10.48550/arXiv.2102.09672},
  \eprint{2102.09672}.

\bibitem[{Oschlies(2021)}]{oschlies_committed_2021}
Oschlies, A., 2021: A committed fourfold increase in ocean oxygen loss.
  \textit{Nature Communications}, \textbf{12~(1)}, 2307,
  \doi{10.1038/s41467-021-22584-4}.

\bibitem[{Oschlies et~al.(2018)Oschlies, Brandt, Stramma,, and
  Schmidtko}]{oschlies_drivers_2018}
Oschlies, A., P.~Brandt, L.~Stramma, and S.~Schmidtko, 2018: Drivers and
  mechanisms of ocean deoxygenation. \textit{Nature Geoscience},
  \textbf{11~(7)}, 467--473, \doi{10.1038/s41561-018-0152-2}.

\bibitem[{Oschlies et~al.(2017)Oschlies, Duteil, Getzlaff, Koeve, Landolfi,,
  and Schmidtko}]{oschlies_patterns_2017}
Oschlies, A., O.~Duteil, J.~Getzlaff, W.~Koeve, A.~Landolfi, and S.~Schmidtko,
  2017: Patterns of deoxygenation: Sensitivity to natural and anthropogenic
  drivers. \textit{Philosophical Transactions of the Royal Society A:
  Mathematical, Physical and Engineering Sciences}, \textbf{375~(2102)},
  20160\,325, \doi{10.1098/rsta.2016.0325}.

\bibitem[{Palter and Trossman(2018)Palter, and
  Trossman}]{palter_sensitivity_2018}
Palter, J.~B., and D.~S. Trossman, 2018: The {{Sensitivity}} of {{Future Ocean
  Oxygen}} to {{Changes}} in {{Ocean Circulation}}. \textit{Global
  Biogeochemical Cycles}, \textbf{32~(5)}, 738--751,
  \doi{10.1002/2017GB005777}.

\bibitem[{Piedagnel et~al.(2025)Piedagnel, Sohail,, and
  Zika}]{piedagnel_mapping_2025}
Piedagnel, E., T.~Sohail, and J.~D. Zika, 2025: Mapping {{Ocean Salinity Data
  Using Gaussian Mixture Modeling}}. \textit{Artificial Intelligence for the
  Earth Systems}, \textbf{4~(3)}, \doi{10.1175/AIES-D-24-0089.1}.

\bibitem[{Price et~al.(2024)}]{price_gencast_2024}
Price, I., and Coauthors, 2024: arXiv:2312.15796. {{GenCast}}:
  {{Diffusion-based}} ensemble forecasting for medium-range weather. arXiv,
  \doi{10.48550/arXiv.2312.15796}, \eprint{2312.15796}.

\bibitem[{Ronneberger et~al.(2015)Ronneberger, Fischer,, and
  Brox}]{ronneberger_unet_2015}
Ronneberger, O., P.~Fischer, and T.~Brox, 2015: arXiv:1505.04597. U-{{Net}}:
  {{Convolutional Networks}} for {{Biomedical Image Segmentation}}. arXiv,
  \doi{10.48550/arXiv.1505.04597}, \eprint{1505.04597}.

\bibitem[{Sall{\'e}e et~al.(2021)}]{sallee_summertime_2021}
Sall{\'e}e, J.-B., and Coauthors, 2021: Summertime increases in upper-ocean
  stratification and mixed-layer depth. \textit{Nature}, \textbf{591~(7851)},
  592--598, \doi{10.1038/s41586-021-03303-x}.

\bibitem[{Sampaio et~al.(2021)Sampaio, Santos, Rosa, Ferreira, P{\"o}rtner,
  Duarte, Levin,, and Rosa}]{sampaio_impacts_2021}
Sampaio, E., C.~Santos, I.~C. Rosa, V.~Ferreira, H.-O. P{\"o}rtner, C.~M.
  Duarte, L.~A. Levin, and R.~Rosa, 2021: Impacts of hypoxic events surpass
  those of future ocean warming and acidification. \textit{Nature Ecology \&
  Evolution}, \textbf{5~(3)}, 311--321, \doi{10.1038/s41559-020-01370-3}.

\bibitem[{Schmidtko et~al.(2017)Schmidtko, Stramma,, and
  Visbeck}]{schmidtko_decline_2017}
Schmidtko, S., L.~Stramma, and M.~Visbeck, 2017: Decline in global oceanic
  oxygen content during the past five decades. \textit{Nature},
  \textbf{542~(7641)}, 335--339, \doi{10.1038/nature21399}.

\bibitem[{Sharp et~al.(2023)Sharp, Fassbender, Carter, Johnson, Schultz,, and
  Dunne}]{sharp_gobaio2_2023}
Sharp, J.~D., A.~J. Fassbender, B.~R. Carter, G.~C. Johnson, C.~Schultz, and
  J.~P. Dunne, 2023: {{GOBAI-O}}{\textsubscript{2}} : Temporally and spatially
  resolved fields of ocean interior dissolved oxygen over nearly 2 decades.
  \textit{Earth System Science Data}, \textbf{15~(10)}, 4481--4518,
  \doi{10.5194/essd-15-4481-2023}.

\bibitem[{Shepherd et~al.(2017)Shepherd, Brewer, Oschlies,, and
  Watson}]{shepherd_ocean_2017}
Shepherd, J.~G., P.~G. Brewer, A.~Oschlies, and A.~J. Watson, 2017: Ocean
  ventilation and deoxygenation in a warming world: Introduction and overview.
  \textit{Philosophical Transactions of the Royal Society A: Mathematical,
  Physical and Engineering Sciences}, \textbf{375~(2102)}, 20170\,240,
  \doi{10.1098/rsta.2017.0240}.

\bibitem[{Shi et~al.(2025)Shi, Jin, Han, Gopalakrishnan,, and
  Narasimhan}]{shi_codicast_2025}
Shi, J., B.~Jin, J.~Han, S.~Gopalakrishnan, and G.~Narasimhan, 2025:
  arXiv:2409.05975. {{CoDiCast}}: {{Conditional Diffusion Model}} for {{Global
  Weather Prediction}} with {{Uncertainty Quantification}}. arXiv,
  \doi{10.48550/arXiv.2409.05975}, \eprint{2409.05975}.

\bibitem[{Shigemitsu et~al.(2017)Shigemitsu, Yamamoto, Oka,, and
  Yamanaka}]{shigemitsu_one_2017}
Shigemitsu, M., A.~Yamamoto, A.~Oka, and Y.~Yamanaka, 2017: One possible
  uncertainty in {{CMIP5}} projections of low-oxygen water volume in the
  {{Eastern Tropical Pacific}}. \textit{Global Biogeochemical Cycles},
  \textbf{31~(5)}, 804--820, \doi{10.1002/2016GB005447}.

\bibitem[{{Sohl-Dickstein} et~al.(2015){Sohl-Dickstein}, Weiss,
  Maheswaranathan,, and Ganguli}]{sohl-dickstein_deep_2015}
{Sohl-Dickstein}, J., E.~A. Weiss, N.~Maheswaranathan, and S.~Ganguli, 2015:
  Deep {{Unsupervised Learning}} using {{Nonequilibrium Thermodynamics}}.
  arXiv, \doi{10.48550/ARXIV.1503.03585}.

\bibitem[{Song et~al.(2022)Song, Meng,, and Ermon}]{song_denoising_2022}
Song, J., C.~Meng, and S.~Ermon, 2022: arXiv:2010.02502. Denoising {{Diffusion
  Implicit Models}}. arXiv, \doi{10.48550/arXiv.2010.02502},
  \eprint{2010.02502}.

\bibitem[{Stramma et~al.(2012)Stramma, Oschlies,, and
  Schmidtko}]{stramma_mismatch_2012}
Stramma, L., A.~Oschlies, and S.~Schmidtko, 2012: Mismatch between observed and
  modeled trends in dissolved upper-ocean oxygen over the last 50 yr.
  \textit{Biogeosciences}, \textbf{9~(10)}, 4045--4057,
  \doi{10.5194/bg-9-4045-2012}.

\bibitem[{Takano et~al.(2023)}]{takano_simulations_2023}
Takano, Y., and Coauthors, 2023: Simulations of ocean deoxygenation in the
  historical era: Insights from forced and coupled models. \textit{Frontiers in
  Marine Science}, \textbf{10}, \doi{10.3389/fmars.2023.1139917}.

\bibitem[{Yoon et~al.(2023)Yoon, Seo, Kim, Choi,, and
  Cho}]{yoon_deterministic_2023}
Yoon, D., M.~Seo, D.~Kim, Y.~Choi, and D.~Cho, 2023: arXiv:2312.02819.
  Deterministic {{Guidance Diffusion Model}} for {{Probabilistic Weather
  Forecasting}}. arXiv, \doi{10.48550/arXiv.2312.02819}, \eprint{2312.02819}.

\bibitem[{Zhang et~al.(2026)Zhang, Ito,, and Bracco}]{zhang_machine_2026}
Zhang, Q., T.~Ito, and A.~Bracco, 2026: A machine learning emulator for
  dissolved oxygen in earth system models. \textit{Environmental Research
  Letters}, \textbf{21~(11)}, 114\,011, \doi{10.1088/1748-9326/ae6fd3}.

\end{thebibliography}

\end{document}